\documentclass{aa}  

\usepackage{graphicx}
\usepackage{txfonts}
\usepackage{threeparttable}
\usepackage{float}
\usepackage{bm}
\usepackage{widetext}
\usepackage{hyperref}

\begin{document} 

  \title{Numerical model of Phobos' motion incorporating the effects of free rotation}
\titlerunning{A numerical model of Phobos' motion incorporating the effects of free rotation.}

  \author{Yongzhang Yang \inst{1,2,7}, Jianguo Yan \inst{2}, Nianchuan Jian \inst{3}, Koji Matsumoto \inst{4,6}, and Jean-Pierre Barriot \inst{2,5}}
  
\authorrunning{Y. Yang et al.}

  \institute{Yunnan Observatories, Chinese Academy of Sciences, Kunming 650216, PR China\\
                              \email{yang.yongzhang@ynao.ac.cn}
              \and
           State Key Laboratory of Information Engineering in Surveying, Mapping and Remote Sensing, Wuhan University, Wuhan 430070, PR China\\
           \email{jgyan@whu.edu.cn}
           \and
             Shanghai Astronomical Observatory, Chinese Academy of Sciences, Shanghai 200030, PR China\\
            \and   
               National Astronomical Observatory of Japan, 2-12 Hoshigaoka, Mizusawa, Oshu, Iwate 023-0861, Japan\\
                             \and
             Observatoire g\'{e}od\'{e}sique de Tahiti, BP 6570, 98702 Faa’a, Tahiti, French Polynesia\\
             \and
             The Graduate University for Advanced Studies, SOKENDAI, Shonan Village, Hayama, Kanagawa 240-0193, Japan\\
             \and
             Key Laboratory of Lunar and Deep Space Exploration, Chinese Academy of Sciences, Beijing 100012, PR China 
       }

   \date{Received Month, Day, Year. accepted Month, Day, Year}  

 
\abstract
  {High-precision ephemerides are not only useful in supporting space missions, but also in investigating the physical nature of celestial bodies. This paper reports an update to the orbit and rotation model of the Martian moon Phobos. In contrast to earlier numerical models, this paper details a dynamical model that fully considers the rotation of Phobos. Here, Phobos' rotation is first described by Euler's rotational equations and integrated simultaneously with the orbital motion equations. We discuss this dynamical model, along with the differences with respect to the model now in use.}
  {This work is aimed at updating the physical model embedded in the ephemerides of Martian moons, considering improvements offered by exploiting high-precision observations
expected from future missions (e.g., Japanese Martian Moons eXploration, MMX), which fully supports future studies of the Martian moons.}
  {The rotational motion of Phobos can be expressed by Euler’s rotational equations and integrated in parallel with the equations of the orbital motion of Phobos around Mars. In order to investigate the differences between the two models, we first reproduced and simulated the dynamical model that is now used in the ephemerides, but based on our own parameters. We then fit the model to the newest Phobos ephemeris published by Institut de M{\'e}canique C{\'e}leste et de Calcul des {\'E}ph{\'e}m{\'e}rides (IMCCE). Based on our derived variational equations, the influence of the gravity field, the Love number, $k_2$, and the rotation behavior were studied by fitting the full model to the simulated simple model. Our revised dynamic model for Phobos was constructed as a general method that can be extended with appropriate corrections (mainly rotation) to systems other than Phobos, such as the Saturn and Jupiter systems.}
  {We present the variational equation for Phobos' rotation employing the symbolic \emph{Maple} computation software. The adjustment test simulations confirm the latitude libration of Phobos, suggesting gravity field coefficients obtained using a shape model and homogeneous density hypothesis should be re-examined in the future in the context of dynamics. Furthermore, the simulations with different $k_2$ values indicate that it is difficult to determine $k_2$ efficiently using the current data.}
   {}

               
   \maketitle
%

\section{Introduction}
\label{sec1}

   The orbital and rotational motions of Phobos have always been of particular interest the keys to understanding the origin and formation of the Martian moons overall \citep{rosenblatt2011origin}. Since its discovery in 1877 and thanks to the development of observation technologies and the first Earth-based and subsequent spacecraft observations, a range of ephemerides, including analytical theories and numerical theories, have been developed \citep{jacobson2014martian}.
   
   One of the first analytical theories for the description of the satellites' orbital motion around Mars was developed by \citet{struve1911uber}. In this theory, Phobos and Deimos are restricted to an elliptical orbit. To explain the satellites' secular perturbation due to the Mars oblateness and the Sun, Struve introduced the precession of pericenter longitude, $\varpi,$ and ascending node, $\Omega$. However, since the perturbation of the orbits are conics, the satellites' semi-major axes and mean motions do not satisfy Kepler's third law very well \citep{jacobson2014martian}. \citet{shor1975the} revised this theory through adding an acceleration term to the mean orbital longitudes to explain the tidal acceleration and estimated the parameters based on observational data collected from 1877 to 1973. Using the variation of arbitrary constants method \citep{brouwer2013methods}, \citet{sinclair1978orbits, sinclair1989orbits} constructed a dynamical theory to obtain an approximate analytical solution to the satellites' equations of orbital motion. The elements $(a, e, \lambda, \varpi, \Omega)$ and the sine of the inclination of the satellite's orbit to the Laplace plane are corrected with secular and periodic terms. Besides the elements, the mean longitude accelerations, the mass of Mars, and the zonal harmonics of the Martian gravity field up to the fourth degree were also included in this theory. To improve the accuracy of the orbit of Phobos and support Soviet Phobos 2 mission, \citet{morley1989catalogue} optimized Sinclair's theory by taking into account more terms including some of second-order perturbations.
   
   Given that the Martian satellites are small, it was found that theories of artificial satellites of the Earth could be adapted as models for their orbits.  Unlike the theories of Struve or Sinclair, those based on the artificial satellite method used the Mars equator, but not the Laplace plane, as the reference plane. With the advent of artificial satellite theory and digital computer technology, a number of analytical and semi-analytical  theories have been developed to describe their orbital motion \citep{ivanov1988theory, kudryavtsev1993calculation, emelyanov1989analytical, emelyanov1993dynamics}.
   
   The most well known and also high-precision semi-analytical approach is Chapront-Touz{\'e} theory \citep{chapront1988esapho, chapront1990orbits, chapront1990phobos}. It takes into account the perturbative effects of the Sun and planets, the Martian gravity field, the mutual perturbations of Phobos and Deimos, the precession and nutation of the Martian equator, and the perturbations due to the gravity field of Phobos. Unlike the orbital elements used in previous theory, the satellite Cartesian coordinates and their velocities in the Mars equator of date coordinate system is first adapted into the model. The most significant improvement of this theory is that it models the libration of Phobos based on the work of \citet{borderies1990phobos}. However, due to the semi-analytical nature of the theory, it is difficult to revise according to new observations or new values for fundamental constants such as the coefficients of Martian and Phobos' gravity field, that is, the numerical terms would have to be regenerated. There have been no updates to the theory published since.
   
   Another approach to studying the motion of Phobos is based on a purely numerical method. The first elegant, purely numerically integrated ephemeris for the Martian satellites was constructed by \citet{lainey2007first} based on their software called Numerical Orbit and Ephemerides (NOE) \citep{lainey2004new2, lainey2004new1}. They modeled the perturbative effects of the Sun, Jupiter, Saturn, Earth, and Moon using DE406 ephemeris, an aspherical Martian gravity field up to the tenth degree and
order  \citep{tyler2003mgs},  mutual perturbations among the two satellites, effects of tides raised on Mars by Phobos \citep{mignard1980evolution},  IAU2000 Martian precession and rotation \citep{seidelmann2002report}, as well as the $C_{20}$ and $C_{22}$ of Phobos. Lainey's integration was implemented in planetocentric Cartesian coordinates referred to the International Celestial Reference Frame (ICRF) with an integrator called RA15 \citep{everhart1985efficient}. They also fitted to observations from 1877 to 2005 and included the spacecraft observations of Mars Global Surveyor (MGS), and Mars EXpress (MEX). The ultimately estimated position accuracy was roughly one kilometer or more with respect to Mars over a century \citep{lainey2007first}. To support the planned Russian Phobos-Grunt mission \citep{marov2004phobos}, \citet{shishov2008determination} employed a similar force model but using Lagrange's equations to integrate the orbital elements. A numerical integration method of the eighth order of accuracy \citep{stepaniants2000effective} was used.
   
   In Lainey's model, they realized that the spin librations of Phobos would greatly affect the accuracy of the model. While the Moon revolves around the Earth in a synchronous rotation state, Phobos is in a synchronous spin-orbit resonance around Mars. Thus, these authors forced Phobos in a $1:1$ spin-orbit resonance directly in their model. \citet{jacobson2010orbits} further developed this model by assuming that Phobos is in synchronous rotation and taking into account only the longitudinal libration (amplitude about $1.24^\circ$ by \citet{willner2014phobos}) during its orbital motion around Mars. It was perfectly reasonable to make such an assumption at the time, since there were no observations of Phobos' Euler angles can be used easily.
   
   Fortunately, this situation will be changed by future missions, which include Martian Moons eXploration (MMX) to be launched in 2024 \citep{kawakatsu2017mission, usui2018martian}. A spacecraft dedicated to the satellite will give us a clearer understanding of the shape and physical characteristics such as gravity field and rotational motion of Phobos in the future.

 To prepare for such future missions, we developed a new pure numerical model of Phobos’ motion that takes no assumption of Phobos' rotation. Phobos' rotational motion is firstly introduced by Euler's rotational equations \citep{cappallo1981numerical, rambaux2012rotational, yang2020elastic}. This full 3D model will enable us to simulate the motion of Phobos with a more realistic case based on the future observation data. To address this issue, we present a revised dynamical model of Phobos' motion around Mars in this paper. Section~\ref{sec2} describes the equations of Phobos' orbital and rotational motion in an appropriate reference system. Section~\ref{sec3} presents the variational equations of the rotational motion. In Section~\ref{sec4}, we use adjustment test simulations to compare the difference between our model and the simple model currently in use.
Finally, we briefly discuss the results and present the main conclusions of this paper in Section~\ref{sec5}.

\section{Dynamical model} 
\label{sec2}
The first step to numerically integrating the motion of Phobos is introducing the modeling process. We naturally divide the modeling process into orbit modeling and rotation modeling.
\subsection{Translational equations of motion}
We developed the equations of motion in a planetocentric (Mars) reference frame with fixed axes align with the ICRF. Hence, in such a reference frame, the position vectors of the Sun and planets relative to Mars can be easily retrieved from the ephemerides. In this paper, we use INPOP19a \citep{fienga2019inpop19a}, the newest version of the Int{\'e}grateur num{\'e}rique plan{\'e}taire de l'Observatoire de Paris (INPOP) planetary ephemerides provided by Institut de M{\'e}canique C{\'e}leste et de Calcul des {\'E}ph{\'e}m{\'e}rides (IMCCE) to get the position and velocity vectors of the Sun and planets relative to Mars in Barycentric coordinates. This ephemeris incorporates updated the Mars observational data by adding the latest data such as ESA’s MEX observations up to 2017, NASA's MGS, Mars Odessey (MO), and Mars Reconnaissance Orbiter (MRO) observations available from 1999 to 2014 \citep{fienga2020asteroid}.

The orbital motion of Phobos around Mars in our model is described by using position ${\bf r} \equiv (\mathrm{r_x}, \mathrm{r_y}, \mathrm{r_z})$ and velocity ${\bf v} \equiv (\mathrm{v_x}, \mathrm{v_y}, \mathrm{v_z})$ in Cartesian coordinates. We can easily write the classical differential equations of relative motion in the planetocentric system as:
\begin{equation}
\label{eq:orb}
\begin{split}
\frac{d^2{\bf r}}{dt^2} = \frac{{\bf F}_\mathrm{p}}{\mathrm{m_p}} - \frac{{\bf F}_\mathrm{0}}{\mathrm{m_0}}\,,
\end{split}
\end{equation}  
 where ${\bf F}_\mathrm{p}$ and ${\bf F}_\mathrm{0}$ indicate the whole external forces exerted on Phobos and Mars, $\mathrm{m_p}$ is the mass of Phobos, $\mathrm{m_0}$ is the mass of Mars, and $t$ is the time expressed in TDB (Barycentric Dynamical Time) timescale.
 
The forces that induce the relative motion can be divided into a two-body force and a perturbation force made up of three parts:\ a part for third-body perturbation, a part for tidal perturbation, and a part for relativistic perturbation. Hence,  $\frac{d^2{\bf r}}{dt^2}$ can be rewritten as :
\begin{equation}
\label{eq:acc}
\begin{split}
\frac{d^2{\bf r}}{dt^2} = {\bf a}_\mathrm{two-body} +  {\bf a}_\mathrm{third-body} +  {\bf a}_\mathrm{tide}  + {\bf a}_\mathrm{rel}\,.
\end{split}
\end{equation} 
 In the barycentric coordinates system, the two-body force contains interaction between Phobos and Mars considered as point mass and interaction between Phobos  and Mars,  taking the form:
\begin{equation}
\label{eq:tb}
\begin{split}
{\bf a}_\mathrm{two-body} =  -\frac{G(\mathrm{m_p} + \mathrm{m_0}){\bf r}}{\mathrm{r^3}} -  G \mathrm{m_0} \nabla \mathrm{U_p} +  G \mathrm{m_0} \nabla \mathrm{U_0}\,,
\end{split}
\end{equation} 
where $G$ is the gravitational constant, $\mathrm{U_p}$ and $\mathrm{U_0}$ denote the gravity potential induced on the punctual body Mars and Phobos, and $\mathrm{r}$ denotes the norm of $\bf{r}$. The gravity potential can be conveniently represented as a spherical harmonic expansion with normalized coefficients ($\bar{C}_{lm}$, $\bar{S}_{lm}$), and is given by \citep{kaula1966theory}:
\begin{equation}
\label{eq:sph}
\begin{split}
 U(r, \phi_{lat}, \lambda) = &\frac{1}{\mathrm{r}} \times \sum_{l=0}^{l_{max}}{\left( \frac{R}{\mathrm{r}} \right)^{l}} \times \\
 &\sum_{m=0}^{l}\bar{P}_{lm}{(\sin\phi_{lat})}[\bar{C}_{lm}\cos m\lambda + \bar{S}_{lm}\sin m\lambda] \,,
 \end{split}
 \end{equation}
where $l$ is the degree, $m$ is the order, $\bar{P}_{lm}$ are the fully normalized Legendre polynomials, $R$ is the reference radius of central body, and ($r, \phi_{lat}, \lambda$)  are the radius, latitude, and longitude in body-fixed coordinates of central body (where $\phi_{lat}$ for the latitude to distinguish from Euler angle $\phi$ below), respectively.

The value of ${\bf a}_\mathrm{two-body}$ in Eq.~\ref{eq:acc} does not refer to an inertial or Newtonian coordinate system, but it is itself subject to an acceleration by the third body \citep{brouwer2013methods}. Hence, denoting the position vector, 
$\bf{r_j}$, of a third-body, $\mathrm{P_j}$, (the Sun, a perturbing planet or Deimos) relative to Mars, and the third-body perturbation can be expressed as 
\begin{equation}
\label{eq:thb}
\begin{split}
{\bf a}_\mathrm{third-body} =  \sum_{j} G \mathrm{m_j} \left( \frac{\bf{r_j - r}}{\mathrm{r^3_{0j}}} - \frac{\bf{r_j }}{\mathrm{r^3_{j}}} \right)\,.
\end{split}
\end{equation} 
Here $\mathrm{m_j}$ is the mass of the third-body $\mathrm{P_j}$, and $\mathrm{r_{0j}}=|\bf{r_j - r}|$. The modeled dynamics included the third-body perturbations due to Deimos, the Earth, Moon, Jupiter, Saturn, and the Sun.

It was \citet{sharpless1945secular} who first found that including an acceleration term can improve his fit to the observed longitudes of Phobos. He concluded that it is a tidal acceleration, but with some uncertainty. Finally, \citet{shor1975the}  found an aptly calculated value when he used new observation data in the fitting and eventually  confirmed the significant influence of the Martian tidal deformation on Phobos' orbital motion.
In principle, the gravitational attraction of both satellites, the planets and the Sun all raise tides on Mars. The elastic response of Mars to these tides is described by Love numbers. The second degree tides are the strongest, and the change in the gravitational potential due to these tides is given by the potential Love number, $k_2$. The tidal potential, $V_{tide}$, caused by a single disturbing body is given by \citep{goossens2008lunar}:
\begin{equation}
\label{eq:stide}
\begin{split}
V_{tide} = \frac{k_2}{2} \frac{G\mathrm{m_{i}}}{R^3_{i}} \frac{a^{5}_{m}}{\mathrm{r^3}} \left( 3(\hat{\bf R}_i \cdot \hat {\bf r})^2 - 1\right)\,,
\end{split}
\end{equation} 
where $\mathrm{m_{i}}$ is the disturbing body,  $R_i$ is the distance between the disturbing body and Mars,  $a_m$ is the equatorial radius of Mars, $\hat{\bf R}_i$ is the unit vector from the centre of Mars to the disturbing body, and $\hat {\bf r}$ is the unit vector from the centre of Mars to observation point (Phobos in this case). 

As for the tide raised by Phobos, instead of Eq.~\ref{eq:stide}, we refer to \citet{lainey2007first} for a full description. Thus, the tidal force ${\bf F}_T$ acting on Phobos takes the form:
\begin{equation}
\label{eq:ptide}
\begin{split}
{\bf F}_T = -\frac{3k_2G\mathrm{m_p}^2a^{5}_{m}}{\mathrm{r^8}} \left(  
{\bf r} +\Delta t \left[
\frac{2{\bf r}({\bf r} \cdot {\bf v})}{\mathrm{r^2}} + ({\bf r} \times \bf \Omega + \bf v)
\right]
\right) \,,
\end{split}
\end{equation} 
where $\bf \Omega$ is the Martian angular velocity vector and $\Delta t$ is the time delay due to the inelastic response of Mars \citep{mignard1980evolution, lainey2007first}. By taking  Eqs.~\ref{eq:stide} and ~\ref{eq:ptide} into account in the calculation of tidal forces on Phobos, the value of ${\bf a}_\mathrm{tide}$ can be modeled.

Finally, we introduce the relativistic effects, ${\bf a}_\mathrm{rel}$. This model is originally inspired by Einstein-Infeld-Hoffmann (EIH) equations of motion used in the planetary and lunar ephemerides \citep{folkner2014planetary, pitjeva2017epm2017, viswanathan2017inpop17a}. For each body (either Mars or Phobos), $A$, the relativistic acceleration due to interaction with other point masses in solar system, ${\bf a}_{A, rel}$ is given by \citep{folkner2014planetary}:
\begin{equation}
\label{eq:rel}
\begin{split}
{\bf a}_{A, rel} &= \sum_{j \neq i} \frac{G\mathrm{m_j}({\bf r}_j - {\bf r}_i)}{r^3_{ij}} 
\bigg \{
-\frac{2(\beta + \gamma)}{c^2}\sum_{k \neq i}\frac{G\mathrm{m_k}}{r_{ik}} - \frac{2\beta -1}{c^2}\frac{G\mathrm{m_k}}{r_{jk}} \\
&+ \gamma \left( \frac{v_i}{c}\right)^2 + (1+\gamma)\left(\frac{v_j}{c}\right)^2 - \frac{2(1+\gamma)}{c^2}{\bf v}_i \cdot {\bf v}_j \\
&-\frac{3}{2c^2}\left[ \frac{({\bf r}_i - {\bf r}_j) \cdot {\bf v}_j}{r_{ij}}\right]^2 + \frac{1}{2c^2}({\bf r}_i - {\bf r}_j)\cdot \frac{d{\bf v}_j}{dt}
\bigg \} \\
&+\frac{1}{c^2}\sum_{j\neq i}\frac{G\mathrm{m_j}}{r^3_{ij}}\left\{({\bf r}_i - {\bf r}_j) \cdot [(2+2\gamma){\bf v}_i - (1+2\gamma){\bf v}_j] \right\}({\bf v}_i - {\bf v}_j)\\
&+\frac{3+4\gamma}{2c^2}\sum_{j \neq i}\frac{G\mathrm{m_j}}{r_{ij}}\frac{d{\bf v}_j}{dt}
\,,
\end{split}
\end{equation} 
where $\beta$ and $\gamma$ are the parametrized post-Newtonian (PPN) parameters and both fixed at 1, with $c$ being the speed of light in vacuum.

Carrying out these formulas Eqs.~\ref{eq:tb}~-~\ref{eq:rel} for ${\bf a}_\mathrm{two-body}$,  ${\bf a}_\mathrm{third-body}$,  ${\bf a}_\mathrm{tide}$, and $ {\bf a}_\mathrm{rel}$ in turn and adding them together, we obtained the orbital motion equations in the planetocentric coordinates.

\subsection{Evolution of Phobos' orientation}
In this paper, Phobos is modeled as an elastic body according to our previous work on Phobos' libration \citep{yang2020elastic}. The orientation of Phobos is integrated from the differential equations for its angular velocities. The angular momentum vectors of Phobos are the product of the angular velocities and the moments of inertia. The angular momentum vectors change with time due to torques and distortion of the body.

To express the variations of Phobos' orientation in the inertial frame, it would be convenient to define Phobos' frame aligned with the principal axes of Phobos. Then the orientation of Phobos' frame with respect to the inertial frame is determined by three Euler angles: $\phi$, $\theta$, and $\psi,$ which evolve over time. The transformation from body-fixed frame (principal axes) to the inertial frame is given by the matrix:
\begin{equation} 
\label{eq:rotm}
\bm{r}_\mathrm{ICRF}= R_{z}(-\phi(t))R_{x}(-\theta(t))R_{z}(-\psi(t))\bm{r}_\mathrm{PA}\,,
\end{equation}
where the rotation matrices, $R_{x}$ and $R_{z}$, are right-handed rotations around the x-axis and z-axis, respectively. Hereinafter, the argument time, $t,$ will be omitted for simplicity.

The instantaneous rate of Euler angle at time, $t$,  $\frac{d \phi}{dt}$, $\frac{d\theta}{dt}$, and $\frac{d\psi}{dt}$, along with the Euler angles, can be used to describe ${\bm{\omega}}(t) \equiv (\omega_1, \omega_2, \omega_3)^{\mathrm{T}}$, the angular velocity of Phobos \citep{goldstein2011classical}:
\begin{equation}
\bm{\omega}
= 
\left[
\begin{matrix}
\sin \theta \sin \psi &\cos \psi&0\\
\sin \theta \cos \psi&- \sin \psi&0\\
\cos \theta&0&1
\end{matrix}
\right]
\left[
\begin{matrix}
&\frac{d \phi}{dt} \\
&\frac{d\theta}{dt} \\
&\frac{d\psi}{dt} 
\end{matrix}
\right]
\,,
\label{eq:eul1}
\end{equation}
where $\mathrm{T}$ means transpose matrix. Then differentiating Eq.~\ref{eq:eul1} with respect to time we obtain a system of equations linear in $\frac{d^2\phi}{dt^2}$, $\frac{d^2\theta}{dt^2}$, and $\frac{d^2\psi}{dt^2}$, for which we algebraically solve, obtaining:
\begin{equation}
\left[
\begin{matrix}
&\frac{d^2 \phi}{dt^2} \\
&\frac{d^2 \theta}{dt^2} \\
&\frac{d^2 \psi}{dt^2} 
\end{matrix}
\right]= 
\left[
\begin{matrix}
\frac{\sin \psi}{\sin \theta}& \frac{\cos \psi}{\sin \theta} &0\\
\cos \psi&- \sin \psi&0\\
0&0&1
\end{matrix}
\right]
\frac{d \bm{\omega}}{dt}
+
\left[
\begin{matrix}
&\frac{\dot \theta (\dot \psi - \dot \phi \cos \theta)}{\sin \theta} \\
&-\dot \phi \dot \psi \sin \theta \\
&-\ddot \phi \cos \theta + \dot \phi \dot \theta \sin \theta
\end{matrix}
\right]\,.
\label{eq:euldd}
\end{equation}
In this paper, the first- and second- order time derivatives of $X$ are also denoted by $\dot X$ and $\ddot X$, respectively.

In a rotating system, the change in angular velocity $\bm{\omega}$ is related to torques $\bf N$ and governed by Euler's rotation equations:
\begin{equation} 
\frac{d}{dt}(\mathbf{I}{\bm \omega}) + \bm{\omega} \times {\mathbf{I}\bm{\omega}} = {\bf N} \,,
\label{eq:eul2}
\end{equation}
where $\mathbf I$ is the moment of inertia tensor. This leads to an equation for $\bm{\omega}$, therefore,
\begin{equation}
\begin{split}
\frac{d{\bm{\omega}} }{dt}= \mathbf{I}^{-1}
\left(
 \mathbf{N}- 
\frac{d\mathbf{I}}{dt} \bm{\omega} - \bm{\omega} \times \mathbf{I} \bm{\omega} 
\right)\,.
\end{split}
\label{eq:eul3}
\end{equation}

\citet{williams2001lunar} first gave the numerical formulas for calculating the inertia tensor of a celestial body with elastic property, these formulas were then implemented to generate the lunar ephemerides \citep{folkner2014planetary, pitjeva2017epm2017} and numerically study the rotation of Phobos \citep{yang2020elastic}.  In this paper, we refer to  \citet{yang2020elastic} for full description to construct the inertia tensor of Phobos. The inertia tensor of the Phobos mainly contains a rigid part $ I_{rigid}$, while it is also subject to tidal distortion from Mars and spin distortion via:
\begin{equation}
\begin{split}
\mathbf{I} = I_{rigid} + I_{tide} + I_{spin} \,.
\end{split}
\label{eq:eul4}
\end{equation}
Here, $I_{tide}$ and $I_{spin}$ are the inertia tensor due to tidal deformation owing to the attraction of Mars and spin distortion, respectively.

The torques, $\bf N,$ referred to the Phobos' frame are generally split into two parts: a torque from point-mass (body) A to the Phobos' figure and a torque from the martian oblateness to the Phobos’ figure:
\begin{equation}
\begin{split}
\mathbf{N} = N_{Pho,fig-pm} +  N_{Pho,fig-fig}  \,,
\end{split}
\label{eq:12}
\end{equation}
The detailed description of $\bf I$ and $\bf N$ can refer to \citet{williams2001lunar}, \citet{folkner2014planetary}, \citet{rambaux2012rotational} and \citet{yang2020elastic}. 

The instantaneous state of Phobos' orientation can be defined completely by six quantities if we choose the Euler angles $(\phi, \theta, \psi)$ and their rates $(\frac{d \phi}{dt}, \frac{d\theta}{dt}, \frac{d\psi}{dt})$ to describe the state of Phobos' orientation, Eqs.~\ref{eq:eul1}, \ref{eq:euldd}, and \ref{eq:eul3} form the differential equations for Phobos' rotational motion.

\subsection{Solution of the equations by numerical integration}
The orbital and rotational equations of motion developed in the previous two sections are non-linear and cross-coupled, and they are not easily solved to determine the position and Euler angles as function of time. Thus, in the implementation of our dynamical model of Phobos' motion, we numerically integrated them simultaneously via Runge-Kutta-Fehlberg (RKF) adaptive procedure method \citep{simos1993runge, yang2018comparison}. Numerical experiments show that the computational procedure is of good accuracy by taking a fixed step size (10 minutes) during the integration.

The RKF integrator is configured for approximating the solution of the first-order differential equation $y' = f(x, y)$ with  an initial condition $y(x_0) = c$, so we rewrite the three second-order orbital equations of motion (Eq.~\ref{eq:orb}) as a set of six first-order differential equations. The result is:
\begin{equation}
\label{eq:orbitalmotion}
\begin{split}
&\frac{d{\bf r}}{dt} = \bf v\\
&\frac{d{\bf v}}{dt} = \frac{d^2{\bf r}}{dt^2} \,.
\end{split}
\end{equation}  
Similarly, the three second-order rotational equations of motion can be rewritten as:
\begin{equation}
\label{eq:rotationalmotion}
\begin{split}
&\frac{d{\Phi}}{dt} = \dot \Phi\\
&\frac{d{\dot \Phi}}{dt} = \frac{d^2{\Phi}}{dt^2} \,,
\end{split}
\end{equation}  
where $\Phi(t) \equiv (\phi, \theta, \psi)^\mathrm{T}$.

The six orbital equations of motion are integrated in parallel with six rotational equations of motion (Eqs.~\ref{eq:orbitalmotion} and \ref{eq:rotationalmotion}). The initial condition state vectors are the three Euler angles and their rates, and position and velocity of Phobos relative to Mars at an initial epoch (J2000.0 in this work). In addition, to control the efficiency and of the numerical integrator and also to validate the computation of the equations of motion and rotation, we use the conservation of energy of the system, that is, neglecte the Martian tidal dissipation.
The full list of physical models and parameters used in the dynamical model is given in Table.~\ref{tbl.1}.
\begin{table*}[!htbp]
\small
\caption{Parameters used in dynamical model.} 
\label{tbl.1}
\centering
\begin{tabular}{lcccccc}
\hline\hline
Parameters && Value&Notes &Reference&\\
\hline
The Sun and planets&&$-$&INPOP19a&\citet{fienga2019inpop19a}&\\
Deimos&&$-$&NOE-4-2020&\citet{lainey2020mars}&\\
&&&&{(ftp://ftp.imcce.fr/pub/ephem/)}\\
Martian gravity field&&$-$&MRO120F, up to degree 12&\citet{konopliv2020detection}&\\
Martian precession+rotation&&$-$&Quoted from MRO120F&\citet{konopliv2020detection}&\\
$k_{2,M}$&&$0.169$&Martian love number&\citet{konopliv2020detection}&\\
$Q_{M}$&&$99.5$&Martian dissipation factor&\citet{jacobson2014martian}&\\
$\delta \bar{J}_{3,M}$&&$-$&Seasonal gravity change of Mars&\citet{konopliv2006global, konopliv2011mars,konopliv2020detection}&\\
Phobos' gravity field&&$-$&Forward model, up to degree 8&\citet{yang2020elastic}&\\
$\mathcal{R}_{Pho}$& &$10.993$ & Radius,km & \citet{willner2014phobos} &\\
$GM_{Pho}$& &$0.7072  \times 10^{-3} $&  $\mathrm{km^3/s^2}$& \citet{patzold2014phobos}&\\
$A,B,C$&&$0.35545, 0.41811, 0.49134$&Moment of inertia, normalized by $MR^2$&\citet{yang2020elastic}&\\
$k_{2,Pho}$&&$2.0 \times 10^{-7}$&Love number of Phobos&\citet{le2013phobos}&\\
\hline
\end{tabular}
\end{table*}
\section{Partial derivatives} 
\label{sec3}
In order to combine our new dynamical model with future higher-precision observations (e.g. MMX) to further the study of Phobos, we rely upon the iterative use of a linear least-squares estimator. Thus, in addition to the model for motion, it is necessary to generate partial derivatives of the state vectors with respect to the rotational and orbital motion initial conditions and parameters we are interested in at all times.

The partial derivatives of the state vectors with respect to initial position and velocity of Phobos and dynamical parameters have been thoroughly studied \citep{peters1981numerical, taylor1998ephemerides, montenbruck2002satellite, lainey2004new2, lainey2004new1}. Here, we focus on the six initial conditions of the rotational state vectors. Denoting the state vectors as $\bm{X}(t) = (\bm{r}(t), \dot{\bm{r}}(t), \Phi(t), \dot{\Phi}(t))^T$, the state transition matrix  could then be obtained from:
\begin{equation}
\label{eq:stm}
\begin{split}
\frac{d}{dt}\left(
\frac{\partial{\bm{X}(t)}}{\partial{\bm{X}(t_0)}}
\right)
=\frac{\partial{\dot{\bm{X}}(t)}}{\partial{\bm{X}(t)}}
\left(
\frac{\partial{\bm{X}(t)}}{\partial{\bm{X}(t_0)}}
\right)\,,
\end{split}
\end{equation} 
and the initial value,
\begin{equation}
\begin{split}
\label{eq:inv}
\left(
\frac{\partial{\bm{X}(t)}}{\partial{\bm{X}(t_0)}}
\right)
\bigg|_{t = t_0}
= \bm{1}_{12 \times 12}\,.
\end{split}
\end{equation} 
We can write $\frac{\partial{\dot{\bm{X}}(t)}}{\partial{\bm{X}(t)}}$ explicitly, 
\begin{widetext}
\begin{equation}
\begin{split}
\label{eq:odesv}
\frac{\partial{\dot{\bm{X}}(t)}}{\partial{\bm{X}(t)}}
=
\left(
\begin{matrix}
\partial{\bm{\dot{r}}}/{\partial\bm{r}} & \partial{\bm{\dot{r}}}/\partial{\bm{\dot{r}}} &\partial{\bm{\dot{r}}}/\partial{{\Phi}} & \partial{\bm{\dot{r}}}/\partial{{\dot{\Phi}}}\\
\partial{\bm{\ddot{r}}}/{\partial\bm{r}} & \partial{\bm{\ddot{r}}}/\partial{\bm{\dot{r}}} &\partial{\bm{\ddot{r}}}/\partial{{\Phi}} & \partial{\bm{\ddot{r}}}/\partial{{\dot{\Phi}}}\\
\partial{{\dot{\Phi}}}/{\partial\bm{r}} & \partial{{\dot{\Phi}}}/\partial{\bm{\dot{r}}} &\partial{{\dot{\Phi}}}/\partial{{\Phi}} & \partial{{\dot{\Phi}}}/\partial{{\dot{\Phi}}}\\
\partial{{\ddot{\Phi}}}/{\partial\bm{r}} & \partial{{\ddot{\Phi}}}/\partial{\bm{\dot{r}}} &\partial{{\ddot{\Phi}}}/\partial{{\Phi}} & \partial{{\ddot{\Phi}}}/\partial{{\dot{\Phi}}}
\end{matrix}
\right)
=
\left(
\begin{matrix}
\bm 0&\bm 1& \bm 0& \bm 0\\
\partial{\bm{\ddot{r}}}/{\partial\bm{r}} & \partial{\bm{\ddot{r}}}/\partial{\bm{\dot{r}}} &\partial{\bm{\ddot{r}}}/\partial{{\Phi}} & \partial{\bm{\ddot{r}}}/\partial{{\dot{\Phi}}}\\
\bm 0 & \bm 0 &\bm 0 & \bm 1\\
\partial{{\ddot{\Phi}}}/{\partial\bm{r}} & \partial{{\ddot{\Phi}}}/\partial{\bm{\dot{r}}} &\partial{{\ddot{\Phi}}}/\partial{{\Phi}} & \partial{{\ddot{\Phi}}}/\partial{{\dot{\Phi}}}
\end{matrix}
\right)
\,.
\end{split}
\end{equation} 
\end{widetext}
To obtain the expression of none-zero components in Eq.~\ref{eq:odesv}, we split them up into four cases:
 \begin{enumerate}
  
  \item $\partial{\bm{\ddot{r}}}/{\partial\bm{r}}$ and $ \partial{\bm{\ddot{r}}}/\partial{\bm{\dot{r}}}$ have been thoroughly studied \citep{peters1981numerical, taylor1998ephemerides, montenbruck2002satellite, lainey2004new2, lainey2004new1} and  were thus omitted here. However, it should be noticed that $\partial{\bm{\ddot{r}}}/{\partial\bm{r}}$ and $ \partial{\bm{\ddot{r}}}/\partial{\bm{\dot{r}}}$ are a  little bit different with previous results since the rotational motion of Phobos were described by Euler angles here. By using Eq.~\ref{eq:rotm}, we get:
 \begin{equation}
\label{eq:pat}
\begin{split}
\bm \xi = R_{z}(\psi)R_{x}(\theta)R_{z}(\phi)\bm{r} \equiv \mathcal {R}(\Phi) \bm{r}\,, 
\end{split}
\end{equation}
where $\bm \xi$ denotes the coordinates referred to the body-fixed (principal axis) system, if we differentiate Eq.~\ref{eq:pat} with respect to time, we obtain the relationship of the velocity in two systems:
\begin{equation}
\label{eq:dpat}
\begin{split}
\frac{d\bm \xi}{dt} = \frac{\partial \mathcal {R}(\Phi)}{\partial t}\bm{r} + \mathcal {R}(\Phi) \dot{\bm{r}}\,.
\end{split}
\end{equation}
Thus,
\begin{equation}
\label{eq:ppat}
\begin{split}
\frac{\partial \bm \xi}{\partial \bm {r}} = \mathcal {R}(\Phi) = \frac{\partial \dot{\bm \xi}}{\partial \dot{\bm {r}}} \,.
\end{split}
\end{equation}   

 \item Based on the Eqs.~\ref{eq:acc} and \ref{eq:tb} above, we are able to conclude that $\partial{\bm{\ddot{r}}}/\partial{{\dot{\Phi}}} = \bm 0$ and $\partial{\bm{\ddot{r}}}/\partial{{\Phi}} =G \mathrm{m_0} \frac{\partial{}}{\partial{\Phi}}  \nabla \mathrm{U_p} $, respectively. To calculate the $\frac{\partial{}}{\partial{\Phi}}  \nabla \mathrm{U_p}$, the inverse transformation from body-fixed frame (principal axes) of Phobos to the inertial frame is needed. This is  discussed in the next section.
 
 \item  To compute $\partial{{\ddot{\Phi}}}/{\partial\bm{r}} $ and $ \partial{{\ddot{\Phi}}}/\partial{\bm{\dot{r}}} $, we differentiate $\ddot{\Phi}$ with respect to the position and velocity of Phobos relative to Mars in inertial frame. Based on Eqs.~\ref{eq:euldd} and \ref{eq:eul3} in this paper, it shows that we only need to evaluate  $\partial{{\dot{\omega}}}/{\partial\bm{r}} $ and $ \partial{{\dot{\omega}}}/\partial{\bm{\dot{r}}} $. This problem is also can be divided into two steps, a transformation from body-fixed frame (principal axes) of Phobos to the inertial frame via the Euler angles and a derivates with respect to the position or velocity, the latter was studied in previous literature.

 \item $\partial{{\ddot{\Phi}}}/\partial{{\Phi}}$ and $\partial{{\ddot{\Phi}}}/\partial{{\dot{\Phi}}}$ will be described in detail in this work.
 
 \end{enumerate}

With the above analysis, we can conclude that the key to establishing the variational equation are $\partial{{\ddot{\Phi}}}/\partial{{\Phi}}$, $\partial{{\ddot{\Phi}}}/\partial{{\dot{\Phi}}}$, and transformation matrix between body-fixed frame (principal axes) of Phobos and the inertial frame. This paper gives these differential equations explicitly.

Referring back to the defining Eq.~\ref{eq:euldd}, we differentiate it for the $\ddot{\Phi}$ with respect to each component of the state vectors:
\begin{equation}
\frac{\partial \ddot \Phi}{\partial \phi}
= 
\left[
\begin{matrix}
\frac{\sin \psi}{\sin \theta} &\frac{\cos \psi}{\sin \theta}& 0\\
\cos \psi &-\sin \psi & 0\\
0&0&1\\
\end{matrix}
\right]
\frac{\partial \bm{\dot \omega}}{\partial \phi}
+
\left[
\begin{matrix}
&0 \\
&0 \\
&\frac{\partial \ddot \phi}{\partial \phi} \cos \theta 
\end{matrix}
\right]
\,,
\label{eq:pdeqn1}
\end{equation}
\begin{equation}
\frac{\partial \ddot \Phi}{\partial \theta}
= 
\left[
\begin{matrix}
\frac{\sin \psi}{\sin \theta} &\frac{\cos \psi}{\sin \theta}& 0\\
\cos \psi &-\sin \psi & 0\\
0&0&1\\
\end{matrix}
\right]
\frac{\partial \bm{\dot \omega}}{\partial \theta}
+
\left[
\begin{matrix}
&(\dot \phi \dot \theta - \ddot \phi) \cot \theta\\
&-\phi \dot{\psi} \cos \theta \\
&\ddot \phi \sin \theta + \cos \theta ( \dot\phi \dot \theta - \frac{\partial \ddot \phi}{\partial \theta})
\end{matrix}
\right]
\,,
\label{eq:pdeqn2}
\end{equation}
\begin{equation}
\frac{\partial \ddot \Phi}{\partial \psi}
= 
\left[
\begin{matrix}
\frac{\sin \psi}{\sin \theta} &\frac{\cos \psi}{\sin \theta}& 0\\
\cos \psi &-\sin \psi & 0\\
0&0&1\\
\end{matrix}
\right]
\frac{\partial \bm{\dot \omega}}{\partial \psi}
+
\left[
\begin{matrix}
&(\dot \omega_1 \cos \psi - \dot \omega_2 \sin \psi)\frac{1}{\sin \theta}\\
&-\dot \omega_1 \sin \psi - \dot \omega_2 \cos \psi \\
&\frac{\partial \ddot \phi}{\partial \psi} \cos \theta 
\end{matrix}
\right]
\,,
\label{eq:pdeqn3}
\end{equation}
\begin{equation}
\frac{\partial \ddot \Phi}{\partial \dot \phi}
= 
\left[
\begin{matrix}
\frac{\sin \psi}{\sin \theta} &\frac{\cos \psi}{\sin \theta}& 0\\
\cos \psi &-\sin \psi & 0\\
0&0&1\\
\end{matrix}
\right]
\frac{\partial \bm{\dot \omega}}{\partial \dot \phi}
+
\left[
\begin{matrix}
&\dot \theta \cot \theta \\
&0 \\
&\dot \theta \sin \theta - \frac{\partial \ddot \phi}{\partial \dot \phi} \cos \theta
\end{matrix}
\right]
\,,
\label{eq:pdeqn4}
\end{equation}
\begin{equation}
\frac{\partial \ddot \Phi}{\partial \dot \theta}
= 
\left[
\begin{matrix}
\frac{\sin \psi}{\sin \theta} &\frac{\cos \psi}{\sin \theta}& 0\\
\cos \psi &-\sin \psi & 0\\
0&0&1\\
\end{matrix}
\right]
\frac{\partial \bm{\dot \omega}}{\partial \dot \theta}
+
\left[
\begin{matrix}
&\frac{1}{\sin \theta}(\dot \psi - \dot \phi \cos \phi) \\
&0 \\
&\dot \phi \sin \theta - \frac{\partial \ddot \phi}{\partial \dot \theta} \cos \theta 
\end{matrix}
\right]
\,,
\label{eq:pdeqn5}
\end{equation}
and
\begin{equation}
\frac{\partial \ddot \Phi}{\partial \dot \psi}
= 
\left[
\begin{matrix}
\frac{\sin \psi}{\sin \theta} &\frac{\cos \psi}{\sin \theta}& 0\\
\cos \psi &-\sin \psi & 0\\
0&0&1\\
\end{matrix}
\right]
\frac{\partial \bm{\dot \omega}}{\partial \dot \psi}
+
\left[
\begin{matrix}
&\frac{1}{\sin \theta} \dot \theta \\
&-\dot \phi \sin \theta \\
& - \frac{\partial \ddot \phi}{\partial \dot \psi} \cos \theta 
\end{matrix}
\right]
\,,
\label{eq:pdeqn6}
\end{equation}
where $\frac{\partial \bm{\dot \omega}}{\partial \phi}$, $\frac{\partial \bm{\dot \omega}}{\partial \theta}$, and $\frac{\partial \bm{\dot \omega}}{\partial \psi}$ are the row elements of Jacobian $\frac{\partial \bm{\dot \omega}}{\partial \Phi}$, and $\frac{\partial \bm{\dot \omega}}{\partial \dot\phi}$, $\frac{\partial \bm{\dot \omega}}{\partial \dot\theta}$, and $\frac{\partial \bm{\dot \omega}}{\partial\dot \psi}$ are the row elements of Jacobian $\frac{\partial \bm{\dot \omega}}{\partial \dot\Phi}$, respectively.

In order obtain the quantities $\frac{\partial \bm{\dot \omega}}{\partial \Phi}$ and $\frac{\partial \bm{\dot \omega}}{\partial \dot\Phi}$, we should again differentiate Eq.~\ref{eq:eul3}, and this time with respect to Euler angles and their rates. We get:
\begin{equation}
\begin{split}
\frac{\partial \dot{\bm{\omega}} }{\partial \Phi}= &\frac{\partial{\mathbf{I}^{-1}}}{\partial \Phi}
\left(
 \mathbf{N}- 
\frac{d\mathbf{I}}{dt} \bm{\omega} - \bm{\omega} \times \mathbf{I} \bm{\omega} 
\right)\\
&+\mathbf{I}^{-1}\left(
\frac{\partial \bm{N}}{\partial \Phi }- \frac{d\mathbf{I}}{dt} \bm{\omega} - \bm{\omega} \times \mathbf{I} \bm{\omega} 
\right)\\
&+\mathbf{I}^{-1}\left( 
\mathbf{N} - \frac{\partial \dot{\bm I}}{\partial \Phi} \bm{\omega} - \dot{\bm I} \frac{\partial \bm{\omega}}{\partial\Phi} - \bm{\omega} \times \mathbf{I} \bm{\omega} 
\right)\\
&+\mathbf{I}^{-1}\left( 
\mathbf{N} -\frac{d\mathbf{I}}{dt} \bm{\omega} -\frac{ \partial \bm{\omega}}{\partial \Phi} \times \mathbf{I} \bm{\omega} -  \bm{\omega} \times \frac{\partial\mathbf{I}}{\partial \Phi}\bm{\omega} - \bm{\omega} \times \mathbf{I} \frac{\partial \bm{\omega} }{\partial \Phi}
\right)
\,,
\end{split}
\label{eq:peul}
\end{equation}
and
\begin{equation}
\begin{split}
\frac{\partial \dot{\bm{\omega}} }{\partial \dot\Phi}= &\mathbf{I}^{-1}\left( 
- \frac{\partial \dot{\bm I}}{\partial \dot \Phi} \bm{\omega} - \dot{\bm I} \frac{\partial \bm{\omega}}{\partial \dot \Phi} - \bm{\omega} \times \mathbf{I} \bm{\omega} 
\right)\\
&+\mathbf{I}^{-1}\left( 
\mathbf{N} -\frac{d\mathbf{I}}{dt} \bm{\omega} -\frac{ \partial \bm{\omega}}{\partial \dot\Phi} \times \mathbf{I} \bm{\omega} -   \bm{\omega} \times \mathbf{I} \frac{\partial \bm{\omega} }{\partial \dot\Phi}
\right)
\,,
\end{split}
\label{eq:pdeul}
\end{equation}
where $\frac{\partial \bm{\omega}}{\partial \Phi}$ and $\frac{\partial \bm{\omega}}{\partial \dot \Phi}$ can be formed directly from Eq.~\ref{eq:eul1}.

In principle any term in Eqs.~\ref{eq:peul} and \ref{eq:pdeul} should be modeled as such to calculate the variational equations, practically, we used \emph{Maple's} \citep{char2013maple} built-in mathematical algorithms for symbolic computation to calculate partial derivatives of the complex formulas, such as $\frac{\partial  \bm{\dot I}}{\partial \Phi}$, and $\frac{\partial{\mathbf{I}^{-1}}}{\partial \Phi}$.

For consistency, with respect to  $\frac{\partial \bm{I}}{\partial \Phi}$, $\frac{\partial  \bm{\dot I}}{\partial \Phi}$, $\frac{\partial{\mathbf{I}^{-1}}}{\partial \Phi}$, and $\frac{\partial \bm{N}}{\partial \Phi }$ expressed in the equations above, we must evaluate the position and velocity components in Phobos' principle axis system. Here,  taking $\frac{\partial \bm{I}}{\partial \Phi}$  as an example, we have:

\begin{equation}
\begin{split}
\frac{\partial \bm{I}}{\partial \Phi} &= \frac{\partial {I_{tide}}}{\partial \Phi} + \frac{\partial {I_{spin}}}{\partial \Phi}\\
&=  \frac{\partial {I_{tide}}}{\partial \bm{\xi}} \frac{\partial \bm{\xi}}{\partial \Phi} + \frac{\partial {I_{spin}}}{\partial \Phi}\\
&= \frac{\partial {I_{tide}}}{\partial \xi_1} \frac{\partial \xi_1}{\partial \Phi} + \frac{\partial {I_{tide}}}{\partial \xi_2} \frac{\partial \xi_2}{\partial \Phi} + \frac{\partial {I_{tide}}}{\partial \xi_3} \frac{\partial \xi_3}{\partial \Phi} + \frac{\partial {I_{spin}}}{\partial \Phi}
\,,
\end{split}
\label{eq:pint}
\end{equation}
where the full descriptions of $I_{tide}$ and $I_{spin}$ can refer to \citet{williams2001lunar}. The terms $\frac{\partial \bm{\xi}}{\partial \Phi}$ represent the change in the principle axis coordinates of a perturbing body (here is Mars) with respect to changes in the Euler angles. 

Carrying out the multiplication for the rotation matrices $R_{z}$ and $ R_{x}$, we obtain the elements of $\mathcal {R}$,
\begin{widetext}
\begin{equation}
\begin{aligned}
\label{eq:rmatrix}
\left[
\begin{matrix}
     \cos \psi \cos \phi- \sin \psi \cos \theta \sin \phi&\cos \psi \sin \phi + \sin \psi \cos\theta \cos \phi & \sin \psi \sin \theta \\
    -\sin \psi \cos \phi - \cos\psi \cos\theta \sin \phi & -\sin \psi \sin \phi + \cos\psi \cos \theta\cos\phi& \cos\psi\sin\theta \\
    \sin \theta\sin\phi&-\sin\theta\sin\phi&\cos \theta
\end{matrix}
\right]\,.
\end{aligned}
\end{equation}
\end{widetext}
Now, $\bm {r}$ does not depend on the orientation of Phobos' principle axis coordinates, so:
\begin{equation}
\label{eq:rpd}
\frac{\partial \bm{\xi}}{\partial{\Phi}} = \frac{\partial \mathcal{R}}{\partial \Phi} \bm{r}
\,.
\end{equation}
Thus, we have
\begin{equation*}
\begin{aligned}
\frac{\partial \xi_1}{\partial \phi} =~&( -\cos \psi \sin \phi - \sin\psi\cos\theta\cos\phi) x \\
&+(\cos\psi \cos\phi -\sin\psi\cos\theta\sin\phi)y
\end{aligned}
\end{equation*}
\begin{equation*}
\begin{aligned}
\frac{\partial \xi_1}{\partial \theta} =~& \sin\psi\sin\theta\sin\phi ~x 
-\sin\psi\sin\theta\cos\phi~y
+\sin\phi\cos\theta ~ z
\end{aligned}
\end{equation*}
\begin{equation*}
\begin{aligned}
\frac{\partial \xi_1}{\partial \psi} =~&( -\sin \psi \cos \phi - \cos\psi\cos\theta\sin\phi) x \\
&-(\sin\psi \sin\phi -\cos\psi\cos\theta\cos\phi)y
+\cos\psi\sin\theta ~ z
\end{aligned}
\end{equation*}
\begin{equation*}
\begin{aligned}
\frac{\partial \xi_2}{\partial \phi} =~&( \sin \psi \sin \phi - \cos\psi\cos\theta\cos\phi) x \\
&-(\sin\psi \cos\phi +\cos\psi\cos\theta\sin\phi)y
\end{aligned}
\end{equation*}
\begin{equation*}
\begin{aligned}
\frac{\partial \xi_2}{\partial \theta} =~& \sin\psi\sin\theta\sin\phi ~x 
-\sin\psi\sin\theta\cos\phi~y
+\sin\phi\cos\theta ~ z
\end{aligned}
\end{equation*}
\begin{equation*}
\begin{aligned}
\frac{\partial \xi_2}{\partial \psi} =~&( -\sin \psi \cos \phi - \cos\psi\cos\theta\sin\phi) x \\
&-(\sin\psi \sin\phi -\cos\psi\cos\theta\cos\phi) y -\sin\psi\sin\theta ~ z
\end{aligned}
\end{equation*}

\begin{equation*}
\begin{aligned}
\frac{\partial \xi_3}{\partial \phi} =~\sin \theta \cos\phi~x - \cos\theta\cos\phi~y
\end{aligned}
\end{equation*}
\begin{equation*}
\begin{aligned}
\frac{\partial \xi_3}{\partial \theta} =~\cos\theta\sin\phi~x -\cos\theta\sin\phi~y-\sin\theta~z
\end{aligned}
\end{equation*}
\begin{equation}
\begin{aligned}
\frac{\partial \xi_3}{\partial \psi} =~0\,.
\end{aligned}
\label{eq:icrs2pa}
\end{equation}

This concludes the derivation of the variational equations for our Phobos' rotational model. For the sake of completeness, we have presented most of the important terms in variational equations for rotational equations of motion. By combining the variational equations for orbital equations of motion and geophysical parameters provided in the previous literatures \citep{lainey2004new1, lainey2007first}, we can build the complete variational equations. In this work, only the positions, velocities, Euler angles, and the rate of Phobos are fitted, while their derivatives are integrated simultaneously with the equations of motion and rotation.

\section{Comparison of the new dynamical model with the current model}
\label{sec4}
The impetus for the refinements to the Phobos' rotational and orbital motion described in this paper are the possible availability of high-precision observations from a future mission, such as trajectory data coming from the probe and the image data of Phobos when the probe is on very close orbits. In addition, a lander and tracking measurement carried out on Phobos would prove especially valuable\citep{kawakatsu2017mission, usui2018martian}. In order to verify the feasibility and effectiveness of the newly constructed model, we need to distinguish the level at which  current models fail as a result of not considering the fully 3D rotation of Phobos. 

\subsection{Reference ephemerides}
\label{sec4.1}

There are a number of parameters in our numerical model whose values affect the orbit significantly, such as the Martian gravity field, Phobos' initial position and velocity, and spin libration of Phobos \citep{lainey2007first}. To reveal the difference between fully coupled approach with the simpler one used so far, 
we first quote the model now employed in the ephemerides \citep{jacobson2010orbits, jacobson2014martian, lainey2020mars}, namely: the simple model.
Since Phobos' rotation period match its orbit period and in synchronous rotation, assuming that the satellite's pole normal to its orbit plane, the angle between the satellite's axis of minimum principal moment of inertia and direction from satellite to Mars is small. In this case, it can be approximated as $\theta = (2e + \mathcal{A})\sin M$, where $e$, $\mathcal{A}$ and $M$ are the satellite's orbital eccentricity, the libration amplitude \citep{rubincam1995gravitational}, and the mean anomaly of satellite in its orbit, respectively. The force on Mars exerted by Phobos based on this assumption can be calculated as:
\begin{equation}
\label{eq:slib}
\begin{split}
{\bf F}_m =  \frac{3}{2}{\mu_p}
\left (
\frac{{a_p}^2}{r^4}
\right )
[(-C_{20} + 6 C_{22} \cos2\theta) ~\hat{\bf{r}} + 4 C_{22} \sin2\theta ~\hat{\bf{t}}] \,,
\end{split}
\end{equation} 
where ${\mu_p}$ is the $GM$ of Phobos, ${a_p}$ is the equatorial radius of Phobos, $C_{20}$ and $C_{22}$ are the second degree harmonic of Phobos, $\hat{\bf{r}}$ is the unit vector directed from Mars toward satellite, and $\hat{\bf{t}}$ is the unit vector in the satellite's orbit plane normal to $\hat{\bf{r}}$ and in the direction of its orbital motion. Easily, the reactive force acting on the satellite is:
\begin{equation}
\label{eq:slib2}
\begin{split}
{\bf F}_s = -
\left(
\frac{\mu_m}{\mu_p}
\right)
~{\bf F}_m
 \,,
\end{split}
\end{equation} 
with $\mu_m$ denoting the $GM$ of Mars.
Using the model presented above, we simulated the current simpler model, but used our own selected physical parameters in Table.~\ref{tbl.1}. Then, we used the six initial conditions (position and velocity) as our adjustment parameters to fit the current ephemeris of Phobos. 
 
\begin{figure}[!htbp]
\centering
\includegraphics[width=\hsize]{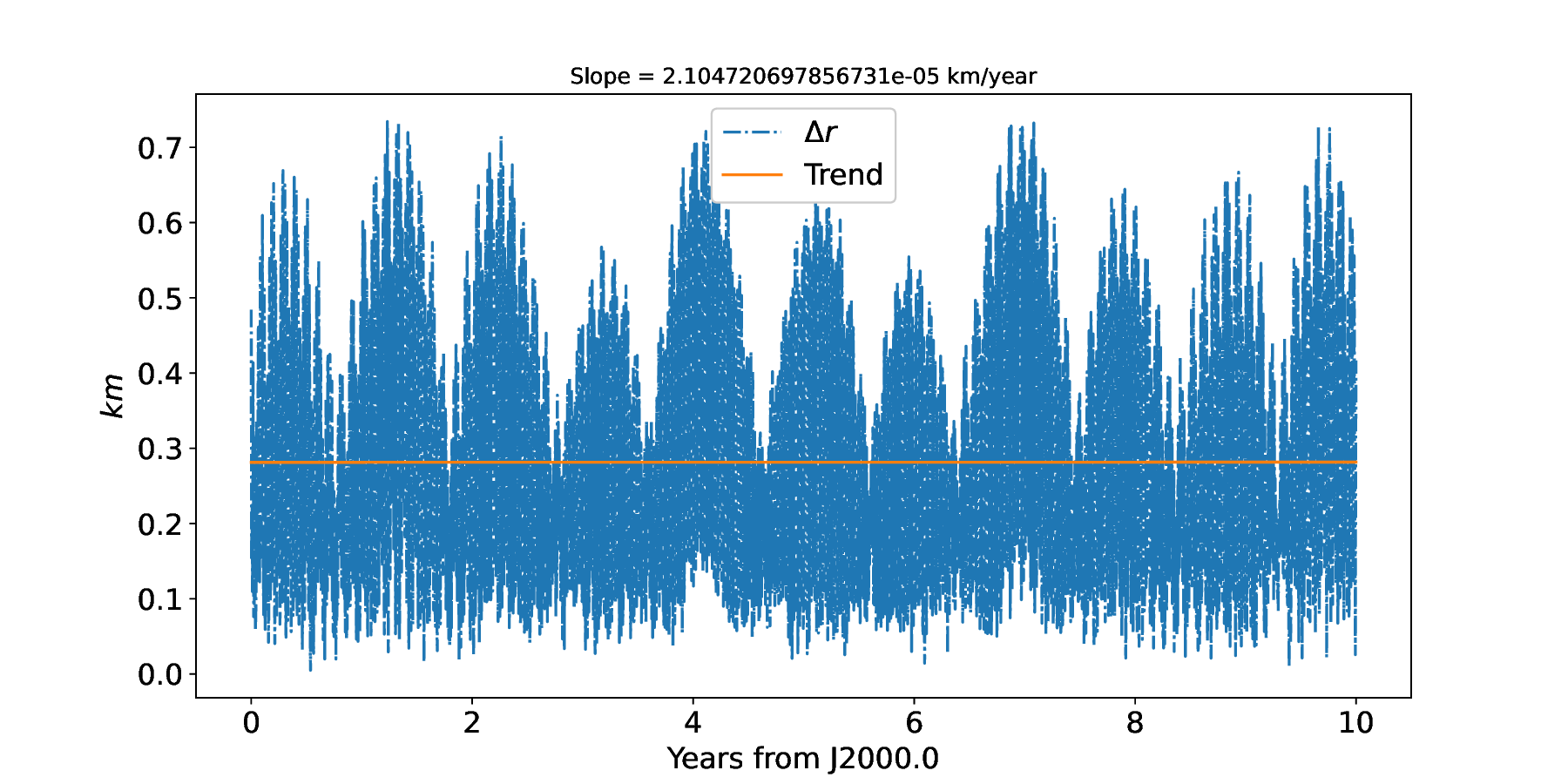}
\caption{Difference in distance after fitting the numerical model to the NOE-4-2020 ephemerides for Phobos.}
\label{orbit_fit}
\end{figure}

We chose the most recent Mars moon ephemerides NOE-4-2020 \citep{lainey2020mars} as the "observations" and fit our model to them. The adjustment was carried out in Cartesian planetocentric coordinates J2000 using a sample of 3650 points with a step size of one day (ten years). We integrated over ten years forth starting at the initial epoch Julian day 2451545.0 (J2000.0, TDB timescale). Figure~\ref{orbit_fit} shows the residuals after applying the least-square fitting procedure. The deviations in the position are probably explained by the different physical parameters (such as the Martian dissipation factor, Q, the physical libration, $\mathcal{A}$, the $C_{20}$ and $C_{22}$ of Phobos, etc.) in these two models. This fitting result gives us an optimal reference for studying the differences between the newly full model and the simpler one used so far.

\subsection{Adjustment to the simpler model}
To explore the difference between our full 3D dynamical model and the previous simple model \citep{jacobson2010orbits, jacobson2014martian, lainey2020mars}. We adjusted the full rotational model to the reference ephemeris integrated in Section ~\ref{sec4.1}. Firstly, the gravity coefficients of Phobos are truncated to the second degree, so as to distinguish the differences that are only due to the modeling.

\begin{figure}[!htbp]
\centering
\includegraphics[width=\hsize]{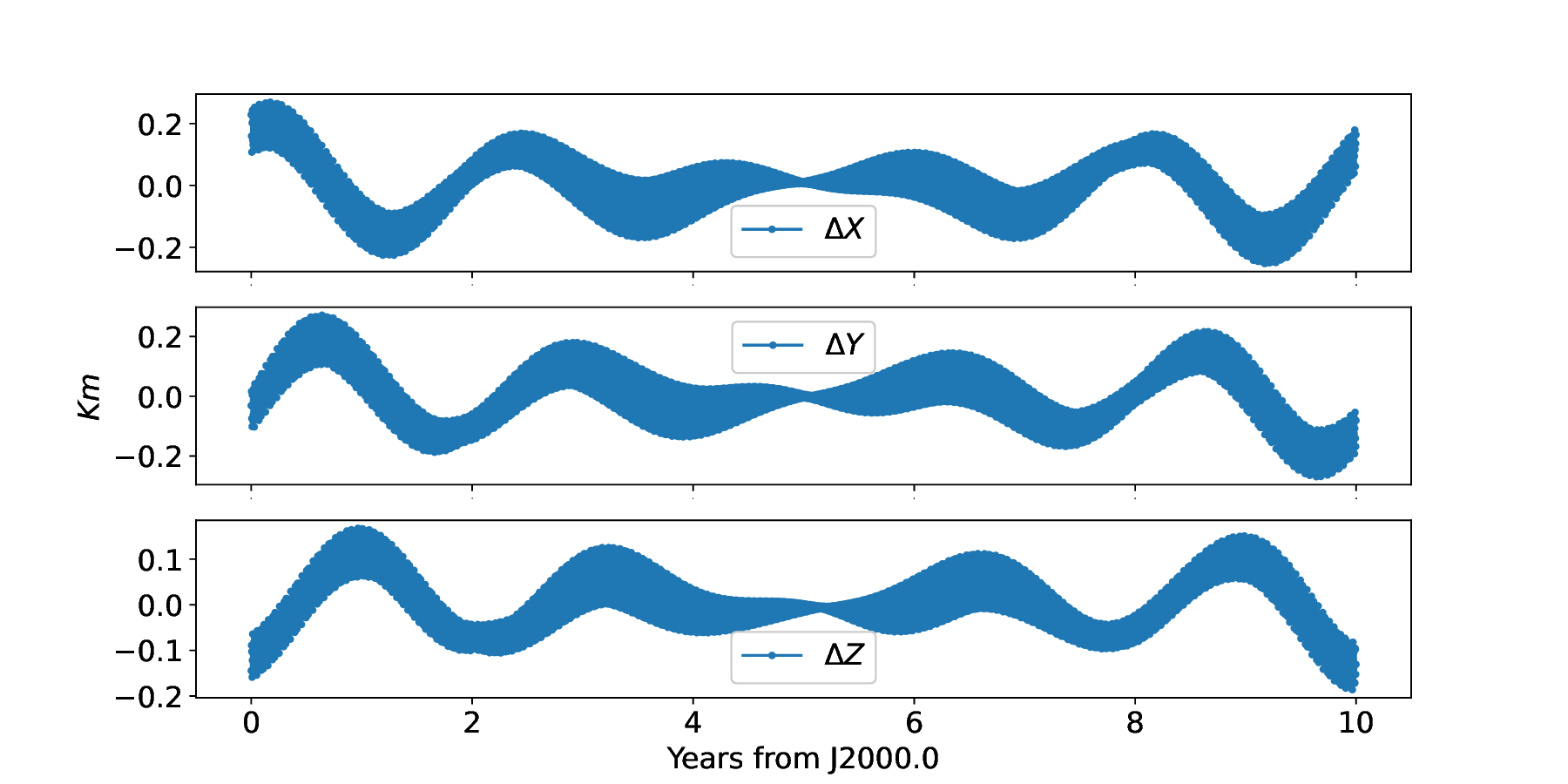}
\caption{Residuals after an adjustment over 10 years of our full model to the simulated simpler model}
\label{2nd_fit}
\end{figure}

\begin{figure}[!htbp]
\centering
\includegraphics[width=\hsize]{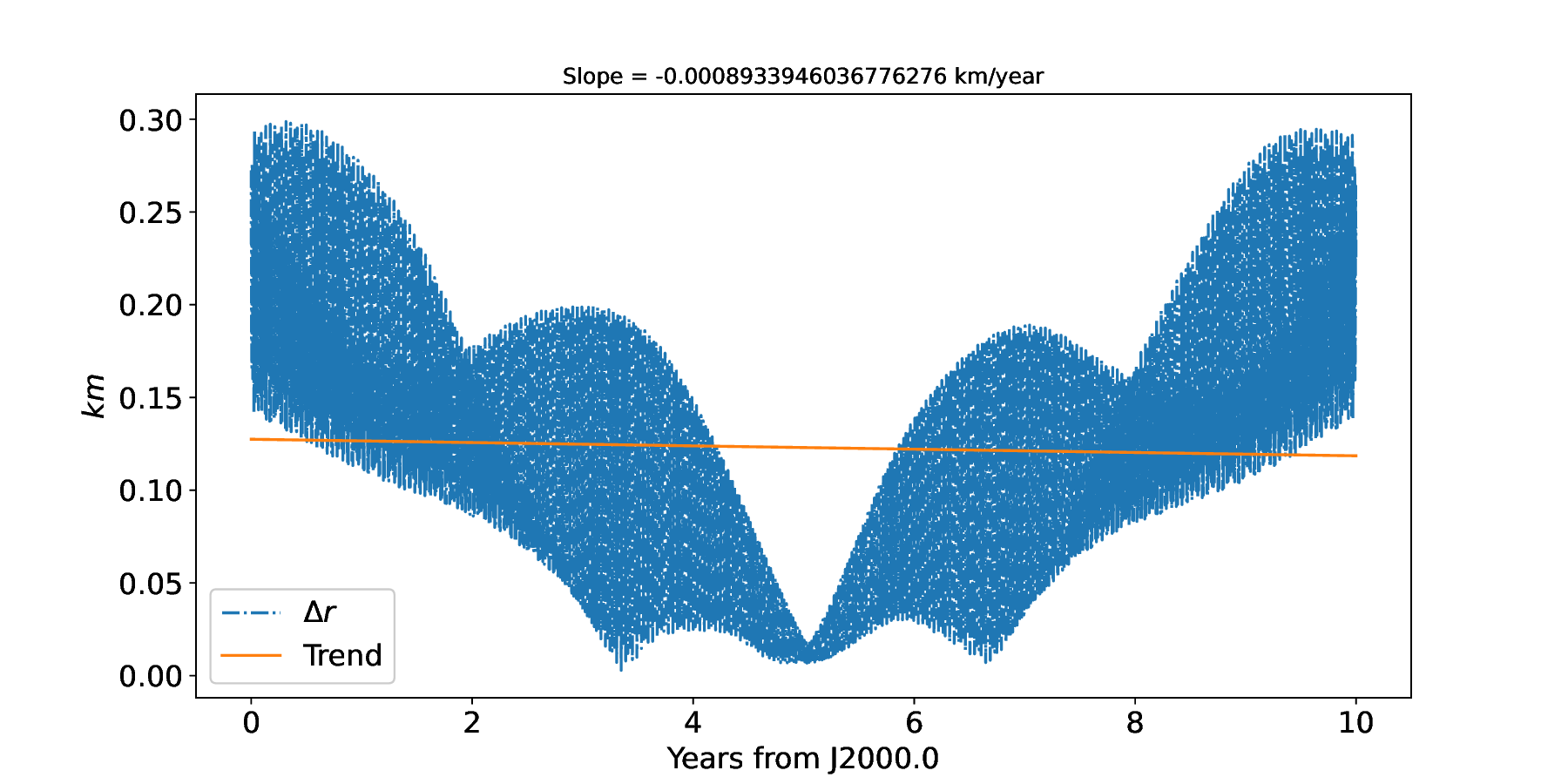}
\caption{Difference in distance after 10 years of fitting the full model to the simulated simpler model.}
\label{dis_fit}
\end{figure}

\begin{figure}[!htbp]
\centering
\includegraphics[width=\hsize]{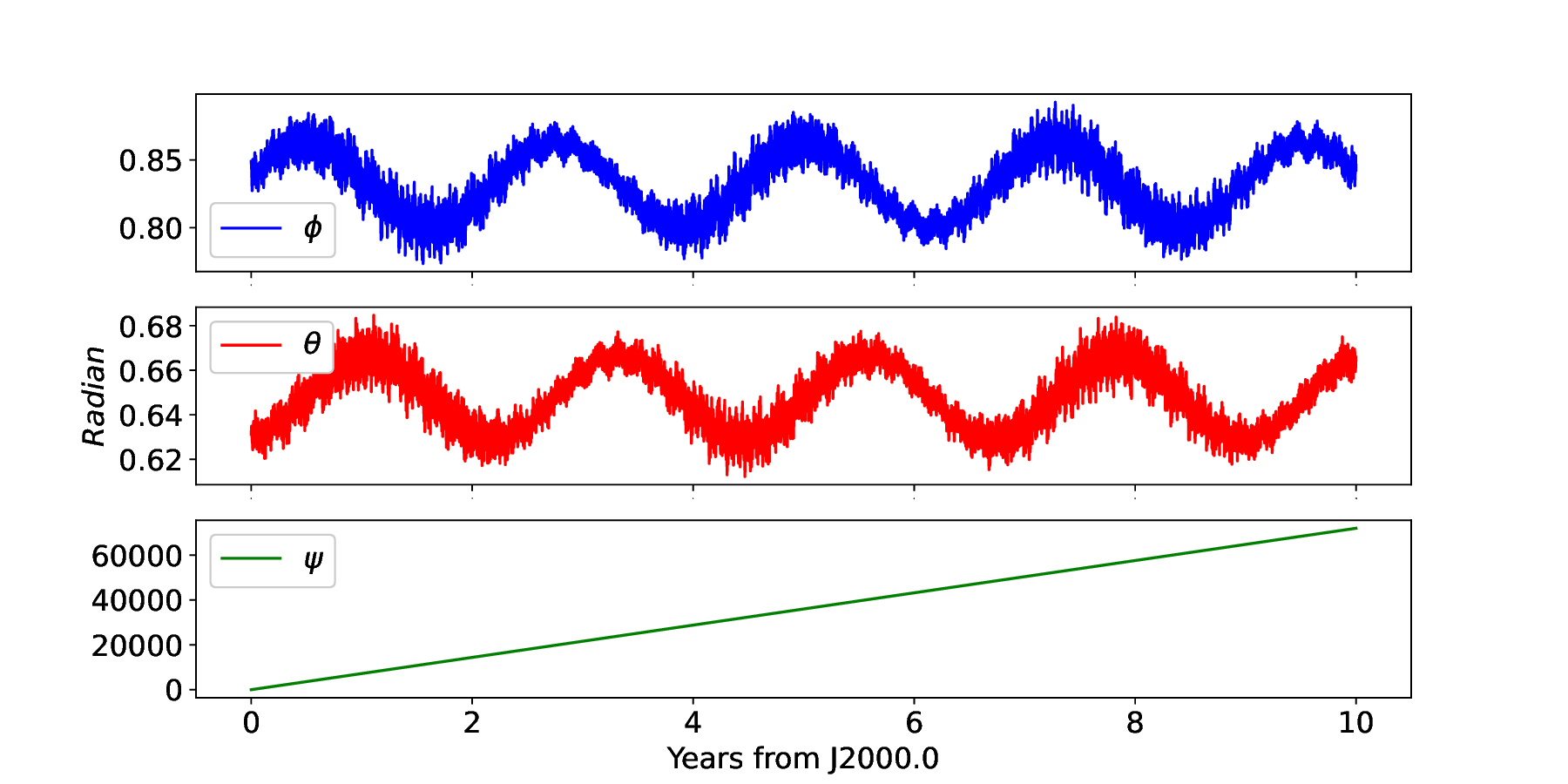}
\caption{Temporal evolution of Phobos' Euler angles about 10 years.}
\label{euler_evol}
\end{figure}

In this paper, we selected the twelve initial conditions (position, velocity, Euler angles, and their rates) as our adjustment parameters to fit the current reference results integrated from the simulated simple model in previous section. The initial position and velocity are retrieved from 
NOE-4-2020 directly. Phobos' Euler angles and its rates refer to \citet{archinal2018report} and the fitting strategy is the same as the previous one. Figures~\ref{2nd_fit} and ~\ref{dis_fit} show the residuals and difference after applying the least-square fitting procedure to our simulated simple model, the deviations of the position are probably explained by the newly introduced latitudinal libration and different longitudinal libration of Phobos in the full model. The evolution of the Euler angles defined respect to the inertial reference frame system for 10 years from 1 Jan 2000 to 2010 are plotted in Fig.~\ref{euler_evol}.

\begin{figure}[!htbp]
\centering
\includegraphics[width=\hsize]{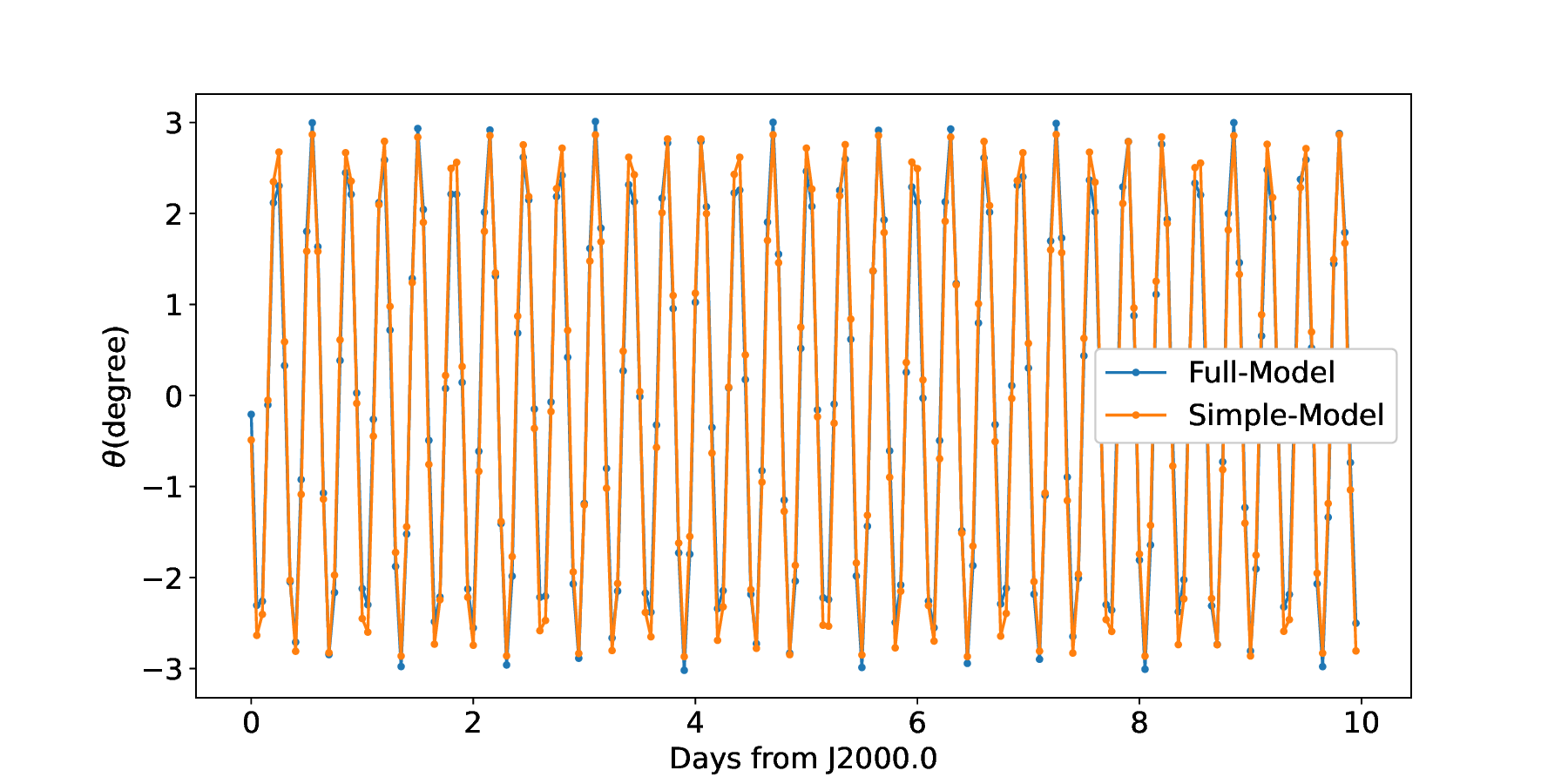}
\caption{Angles between the direction from Phobos to Mars and Phobos’s axis of minimum principal moment of inertia.}
\label{lon_osc}
\end{figure}

\begin{figure}[!htbp]
\centering
\includegraphics[width=\hsize]{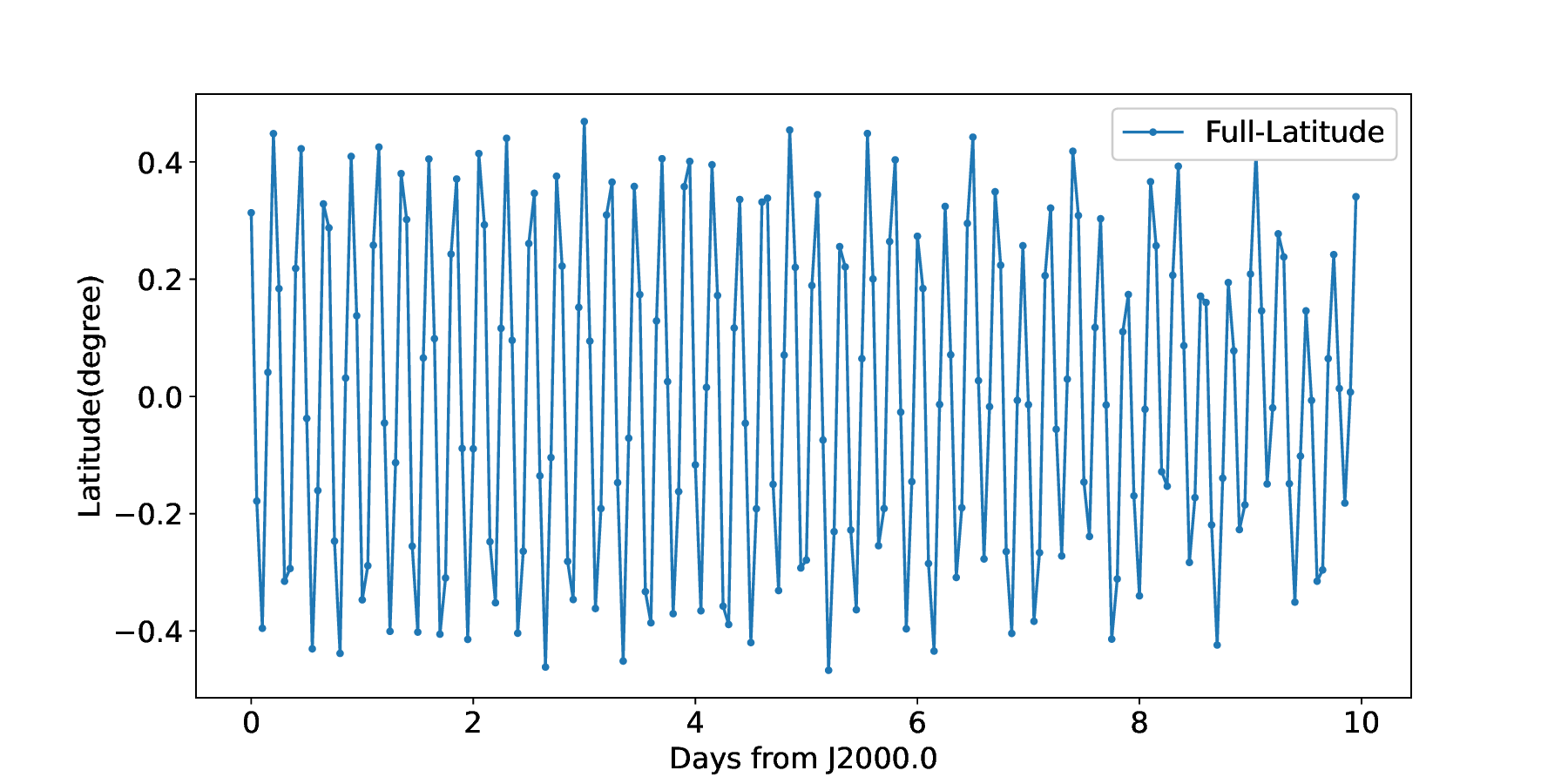}
\caption{Angles between the pole of Phobos and normal of Phobos’s orbital plane.}
\label{lat_osc}
\end{figure}

\begin{figure}[!htbp]
\centering
\includegraphics[width=\hsize]{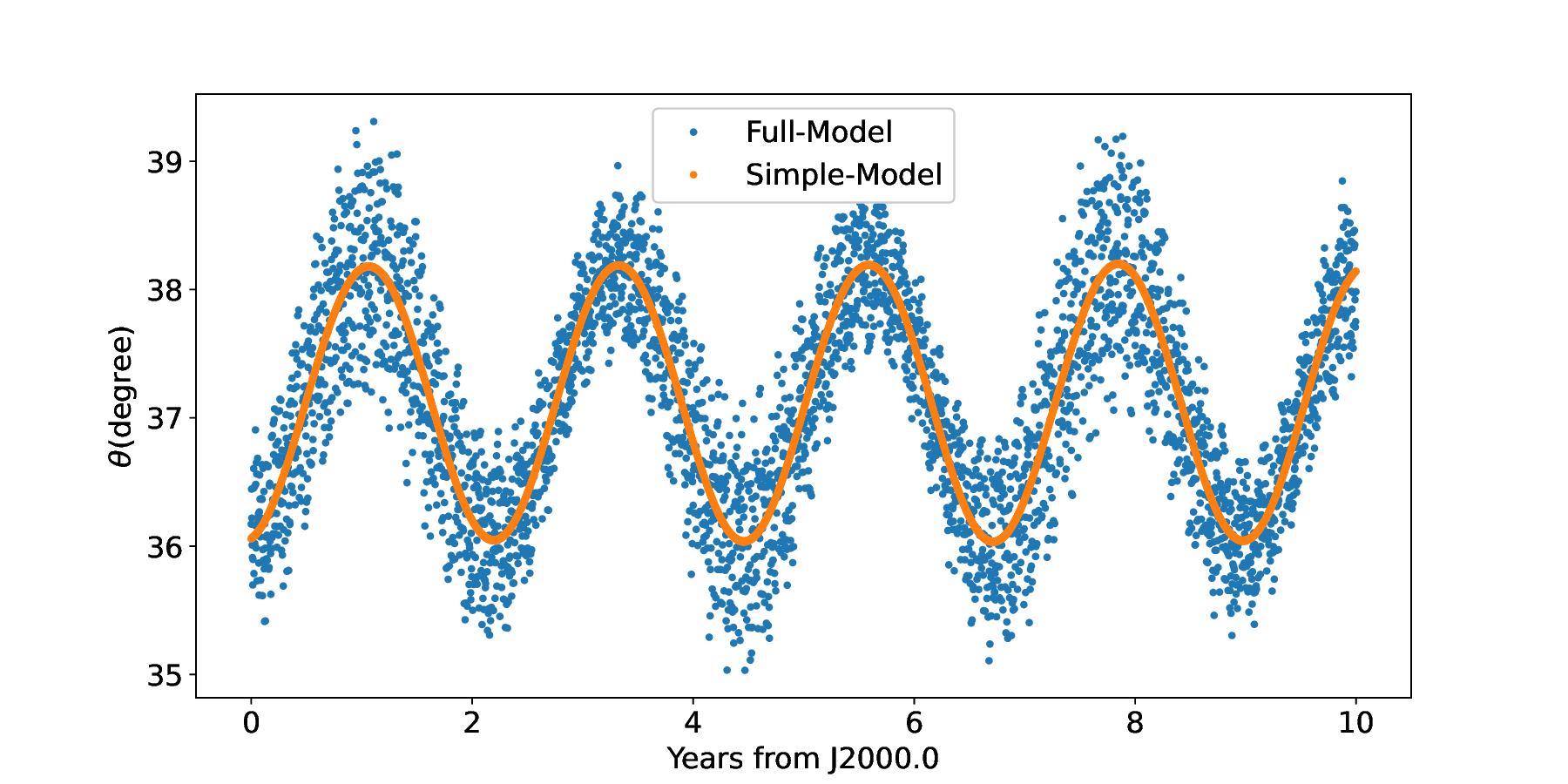}
\caption{Temporal evolution of Phobos' pole position on inertial space about 10 years.}
\label{pole_osc}
\end{figure}

To analyze the rotation behavior, Fig.~\ref{lon_osc} presents the angles between the direction from Phobos to Mars and Phobos's axis of minimum principal moment of inertia in the two models, respectively. The differences between the two dynamical models are very small, indicating that the simple model can describe the deviation in the longitude direction very well. The Phobos obliquity is shown in Fig.~\ref{lat_osc}, this quantity is neglected in the simple model, so this may be a reason for the different orbits, even though it is an order of magnitude smaller than the angle in the longitude direction. The difference between the two models are then the instantaneous spin pole on the space. On average, the two poles are moving on the same rate that is shown and confirmed in Fig.~\ref{pole_osc}. However, the thickness of the blue points in Fig.~\ref{pole_osc} indicate the full model has a large oscillation amplitude of around one degree, compared to the simple model which assumes the Phobos pole is aligned with its orbit normal.

\subsection{Full model with higher gravity}

The other main purpose of this work is to examine the effect of introducing higher order gravity field coefficients on rotational and orbital motions. However, because the gravity field of Phobos is hard to determine directly from the current data, especially as the coefficients higher than two degrees are now all derived from the shape of Phobos with a homogeneous density hypothesis. Table.~\ref{tbl.2} provides the numerical values of these coefficients up to the fourth degree and order from a forward modeling method\citep{yang2020elastic}.

\begin{table}
\caption{Gravity coefficients for a homogeneous Phobos obtained with the forward modeling method from \citet{yang2020elastic}.The reference radius is $R = 14.0\ km$.}             
\label{tbl.2}      
\centering                          
\begin{tabular}{c c c c}        
\hline\hline                 
$l$& $m$ & $\bar{C}_{lm}$ & $\bar{S}_{lm}$ \\    
\hline                        
2&      0&      -0.029395399889203 &    0.000000000000000 \\
2&      1&      -0.000000000001687 &    0.000000000000989 \\
2&      2&      0.015254365665109 &     -0.000000000007499 \\
3&      0&      0.001502237000330 &     0.000000000000000 \\
3&      1&      0.002073579621118        &    -0.001024791299026 \\
3&      2&      -0.004400837773116 &    0.000523900403235 \\
3&      3&      -0.000593703652260 &    0.006612763731665 \\
4&      0&      0.002558008160322 &     0.000000000000000 \\
4&      1&      -0.001340635723883 &    0.000402028977734 \\
4&      2&      -0.000924218398140 &    -0.000632299744182 \\
4&      3&      0.001239020990099 &     -0.001058810893800 \\
4&      4&      0.000326748849574 &     0.000029213145563 \\
\hline                                   
\end{tabular}
\end{table}

\begin{figure}[!htbp]
\centering
\includegraphics[width=\hsize]{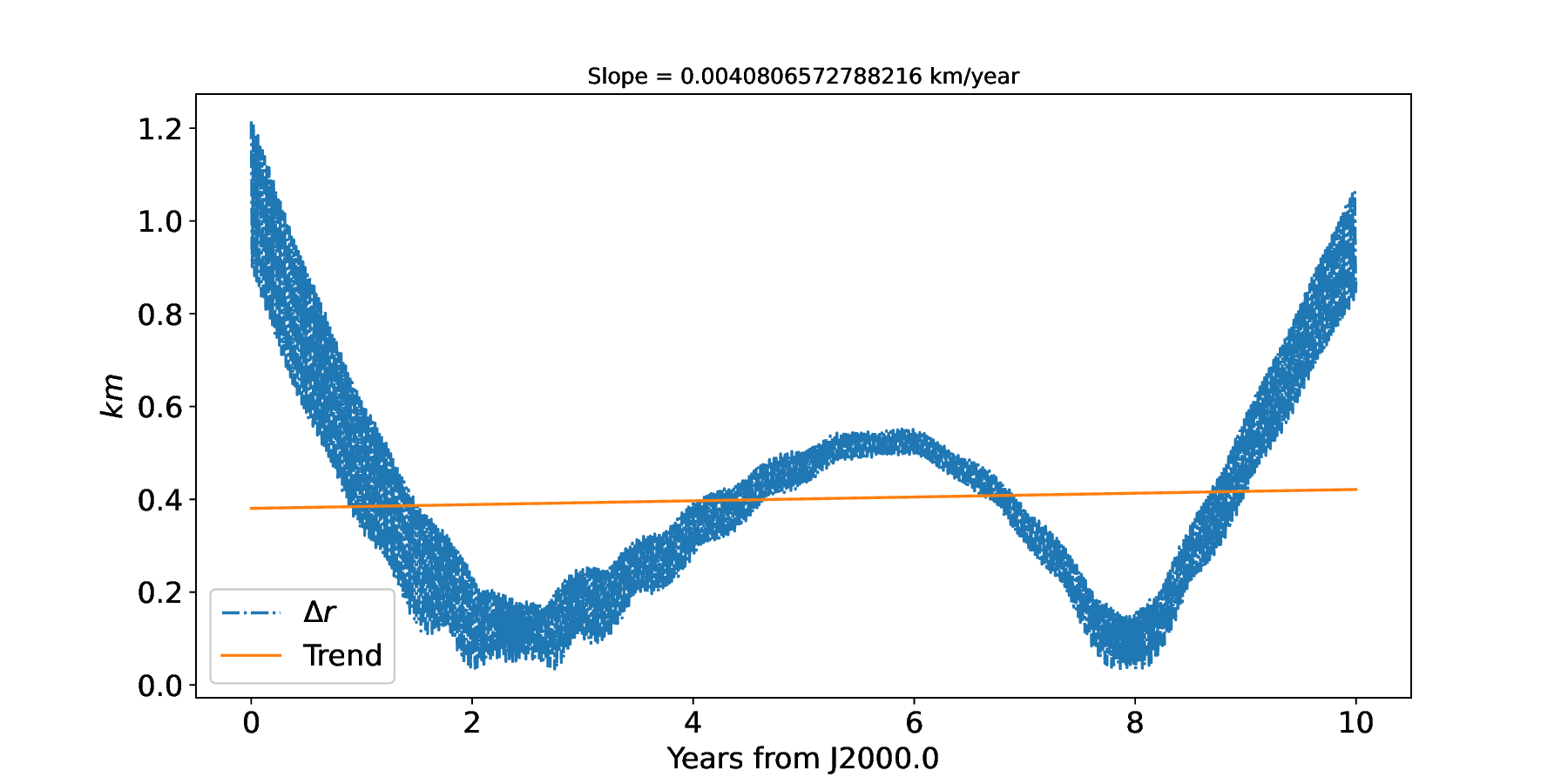}
\caption{Difference in distance after 10 years of fitting the full model with gravity filed truncated at the third degree to the simulated simpler model.}
\label{dis_d3}
\end{figure}

\begin{figure}[!htbp]
\centering
\includegraphics[width=\hsize]{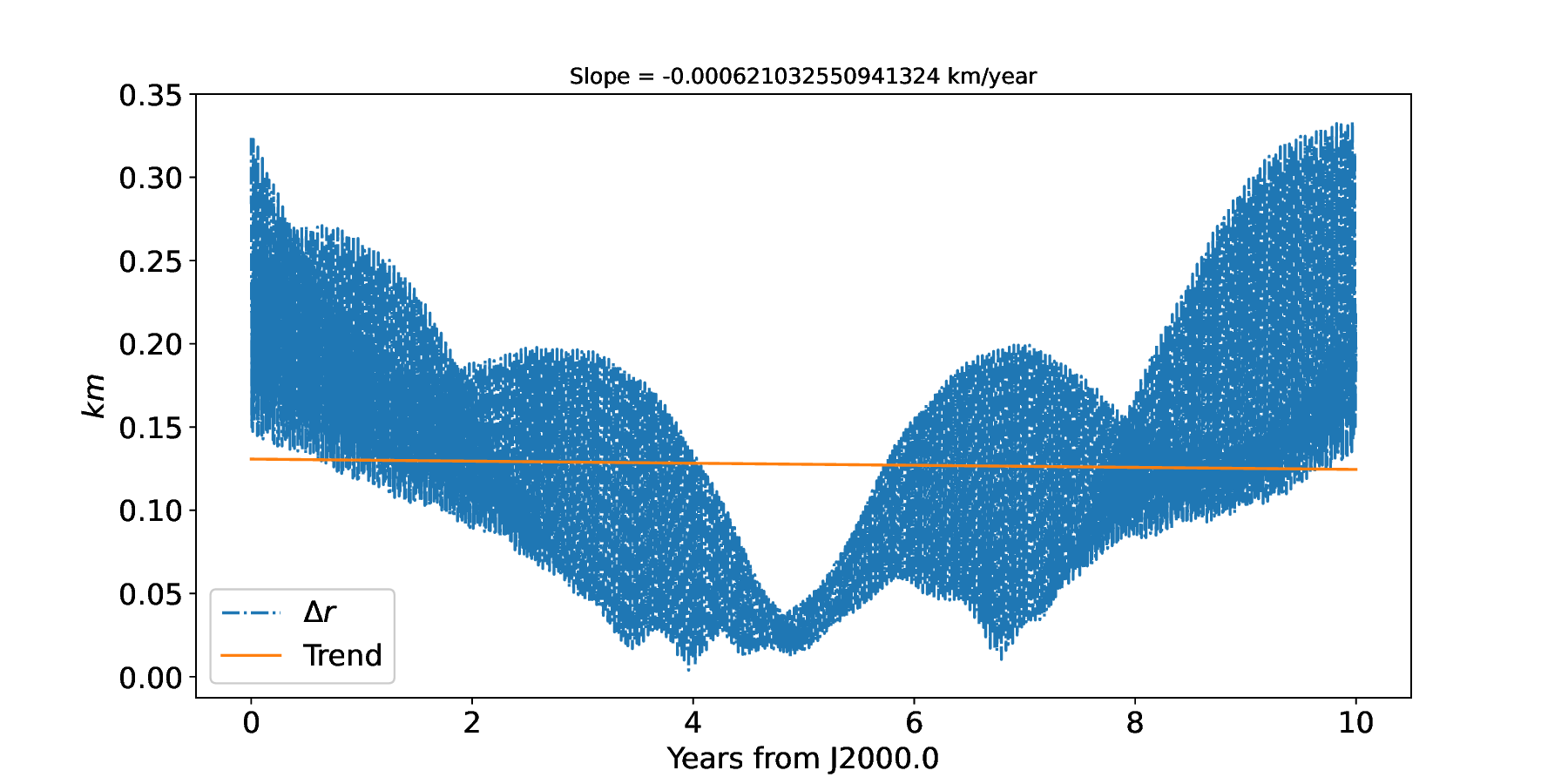}
\caption{Difference in distance after 10 years of fitting the full model with gravity filed truncated at the third degree but without $S_{33}$ to the simulated simpler model.}
\label{3rd_ns33}
\end{figure}

Figure~\ref{dis_d3} displays the difference after an adjustment about 10 years of our full model with Phobos' gravity truncated at a degree and order of three to the simulated simple model in Sect.~\ref{sec4.1}. One can wonder why the difference in distance after fitting the model that takes into account the third-degree gravity field of Phobos to simulated simple model are more significant than that of a model that takes into account only the second degree. During our adjustment we found $S_{33}$ to be the cause of large orbit difference, that is, if we neglect the $S_{33}$ and the adjustment will be very close to the case where only the second-degree gravity field is considered. Fig.~\ref{3rd_ns33} presents the difference in distance after fitting the full model with the third-degree gravity field, but not $S_{33}$ to the simulated simple model of Phobos. The resulting difference is very close to the full model of Phobos with gravity coefficients truncated at the second degree. This result may come from the newly introduced librations by the third-degree harmonic \citep{borderies1990phobos, rambaux2012rotational}, especially with respect to the relatively large value of $S_{33}$, and the tidal orbital expansion may also bring about this signal \citep{lainey2020resonance}. In conclusion, this issue needs to be carefully re-studied in the future based on more reliable gravity field coefficients of Phobos.

\begin{figure}[!htbp]
\centering
\includegraphics[width=\hsize]{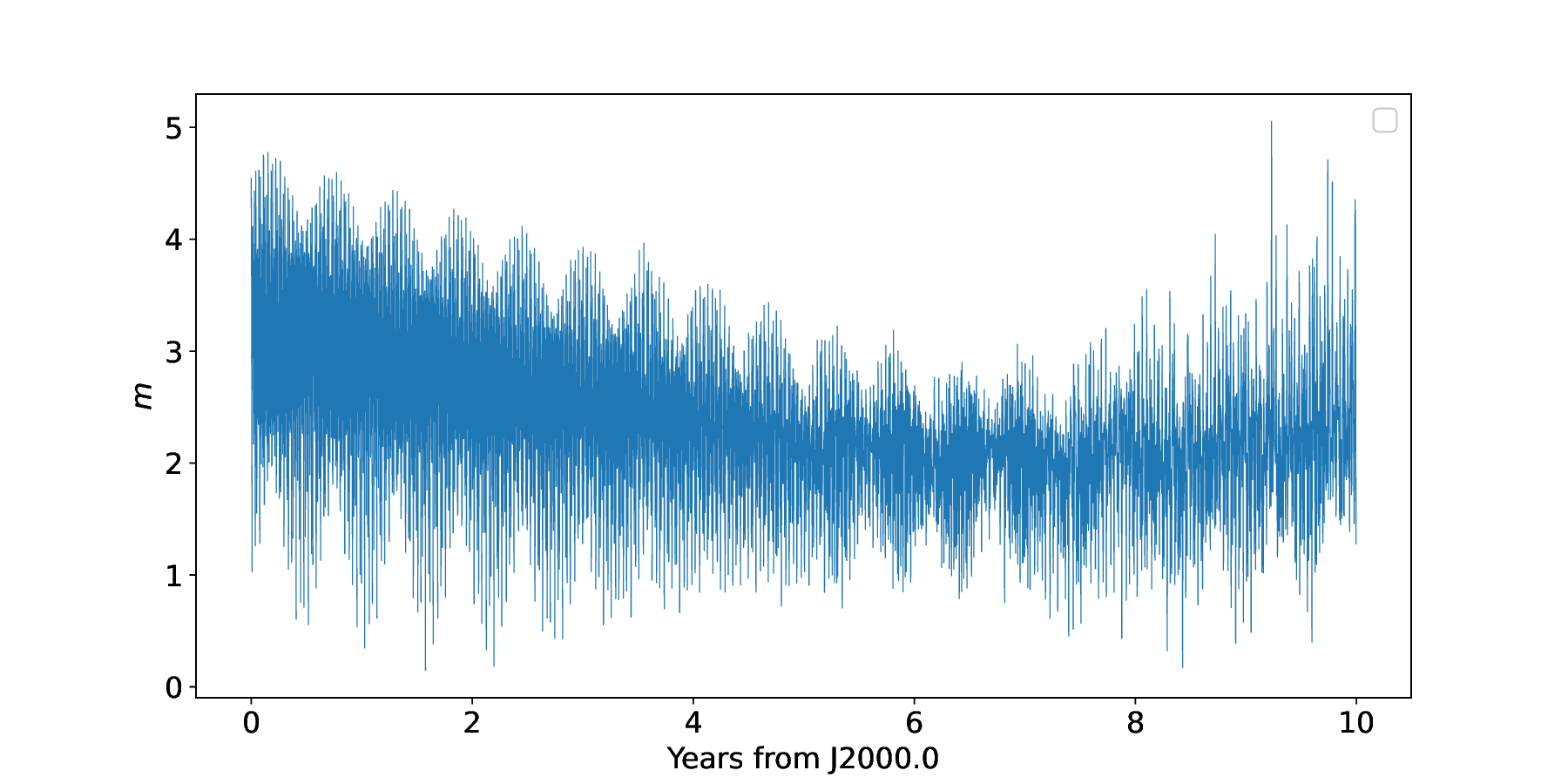}
\caption{Difference in distance after 10 years of fitting the full model with gravity filed truncated at the fourth and third degrees to the simulated simpler model. Note: the unit of the $Y$-axis is meters.}
\label{deg_34}
\end{figure}

\begin{figure}[!htbp]
\centering
\includegraphics[width=\hsize]{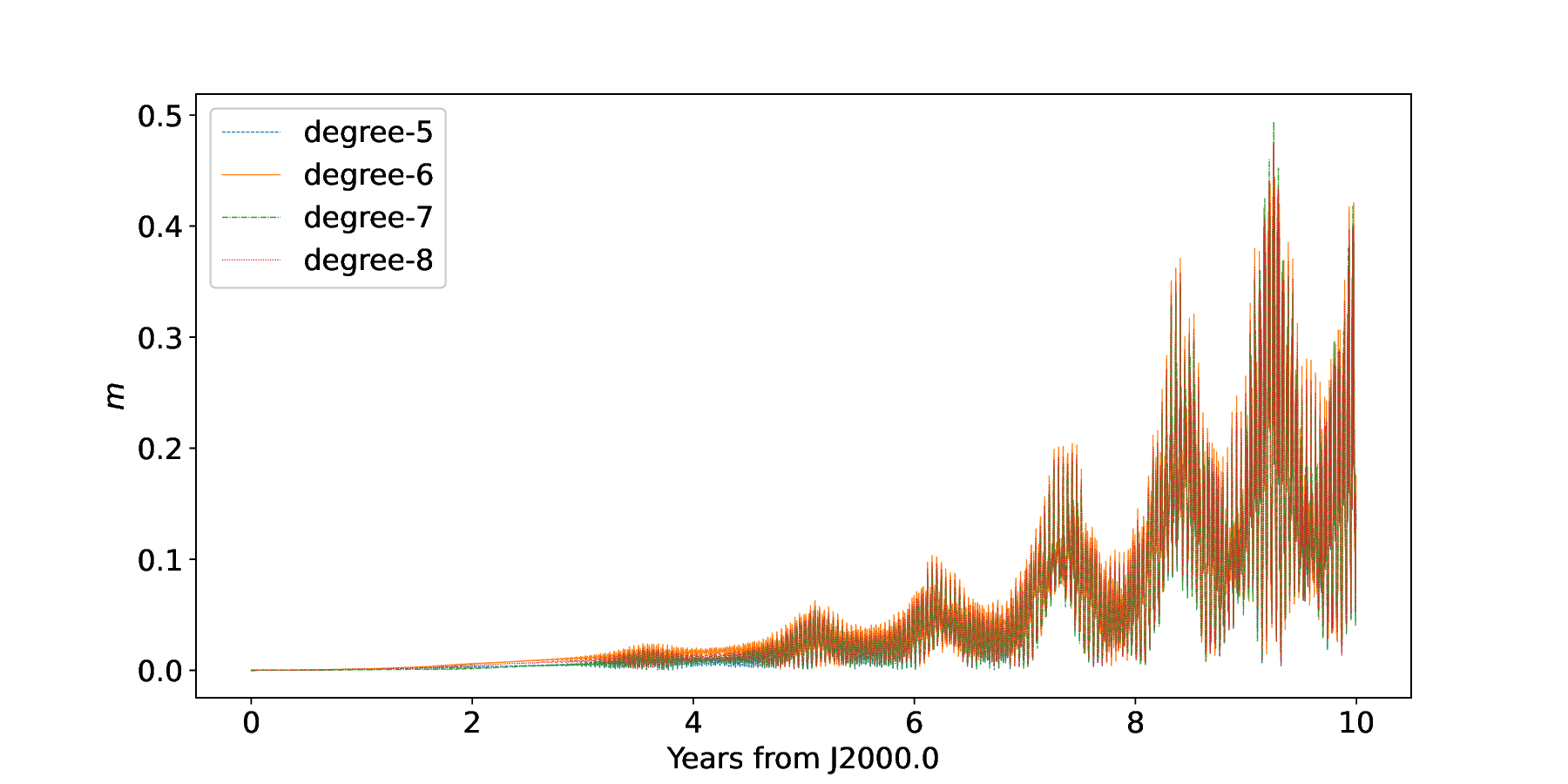}
\caption{Difference between the integration results of the full model using the higher order gravity field and the fourth-degree model, with the initial values all using the fitted values of the fourth-degree model. Note: the unit of the $Y$-axis is meters.}
\label{deg_4plus}
\end{figure}

We compared the results of fitting the full model considering higher degree $(i = 3, \cdots, 8)$ gravity field coefficients to the simulated simple model.
The difference of the post-fitted orbits containing third- and fourth-order gravity field coefficients is plotted in Fig.~\ref{deg_34}, it indicates that the difference between the two simulations is at the "meter" level. In addition, we find that the maximum orbit difference does not exceed one meter$~$ even when directly integrating a model containing higher-order gravity field coefficients using the post-fitted initials of a model containing fourth-order gravity field coefficients, as shown in Fig.~\ref{deg_4plus}. 
This suggests that the degree of gravity field coefficients higher than three can be neglected in the full model concerning the current accuracy. However, this conclusion is strongly dependent on observational data and will need to be revisited in the future if lander tracking data on Phobos' surface or high-precision rotation data become available.

\subsection{Effect of Phobos' tidal Love number, $k_2$}
\label{k2}
Some uncertainty about the elastic property of Phobos remains. In this work, Phobos was treated as an elastic body where the tidal Love number, $k_2$, was referred to by \citet{le2013phobos} to construct the numerical model of motion. The value of $k_2$ affects the rotation of Phobos \citep{williams2001lunar, yang2020elastic} and, thus, the orbital motion \citep{lainey2007first, jacobson2010orbits, jacobson2014martian}.
Consequently, in order to identify the effect of different $k_2$ on Phobos’ orbital motion, we chose various $k_2$ to simulate Phobos' motion. 

\begin{figure}[!htbp]
\centering
\includegraphics[width=\hsize]{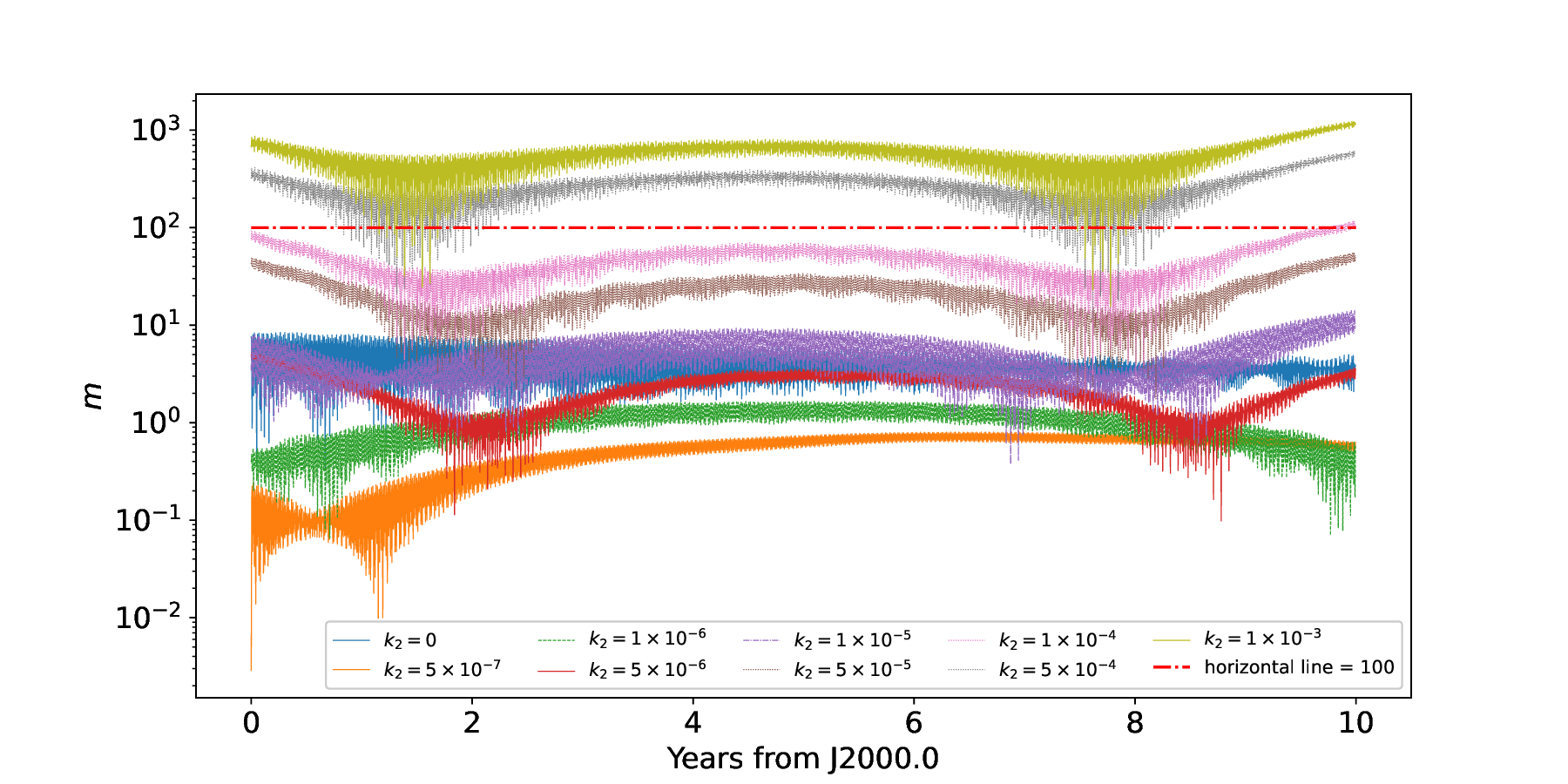}
\caption{Post-fitted orbital position differences over 10 years between using different $k_2$ with $k_2 = 1.0 \times 10^{-4}$. Note: the unit of the $Y$-axis is meters.}
\label{k2_var}
\end{figure}

By including a range of plausible value for $k_2 \in [1.83\times 10^{-7}, 5.3\times 10^{-4}]$ \citep{le2013phobos}, our new model uses a second-degree gravity field of Phobos and a different $k_2$ value fitted to the simulated simple model that without considering $k_2$ and then the differences (with and without)  are plotted in Fig.~\ref{k2_var}. As shown in this figure, the differences in orbit are less than $100~m$ in ten years' integration when $k_2 \leq 1.0\times 10^{-4}$. This result suggests that it is difficult to effectively determine the $k_2$ value of Phobos from the current data.

\section{Discussion and conclusion} 
\label{sec5}
A high-precision numerical ephemeris can typically provide the accurate positions, velocities, and orientations of celestial bodies over time. Therefore, it can allow for the detailed study of the evolution and internal structure of these celestial bodies. In this paper, we establish a numerical dynamical model of Phobos' motion with full consideration of its rotation. The full "3D" dynamical model was constructed in planetocentric Cartesian coordinates based on integrating Newton's equations for orbital motion and Euler's rotational equations for Phobos' rotation simultaneously. We explicitly gave the variational equations of the rotational motion of Phobos when adjusting the numerical solution to the observations.
In order to exclude the influence of other parameters in the model, we first simulated the simple model currently in use and then fitted it to the current French ephemerides NOE-4-2020,  using the least-squares approach to determine the initial conditions.

We fit full rotation model to the simulated simple model, with the gravity field truncated at only $C_{20}$ and $C_{22}$. The results of the simple and full models are close, but the full model includes more latitudinal motion in the rotational motion. However, when we fit the full model including the third-order gravity field coefficients to the simulated simple model, the results show a large discrepancy, this may be due to the introduction of new librations resulting from the use of gravity field coefficients containing higher orders, a phenomenon that needs to be re-examined in the future based on more plausible gravity field coefficients.

Given the internal structure and elastic property of Phobos have not been clearly determined so far, we introduced the tidal Love number, $k_2$, of Phobos into our model to check its influence on Phobos' motion. The simulations indicate that this perturbation is non-negligible when the $k_2$ value is larger than $1.0\times 10^{-4}$. Otherwise, the satellite can be treated as a rigid body when the $k_2$ is no larger than $1.0\times 10^{-4}$, which will make the integration and fit process more efficient and easier.

This revised numerical model of Phobos' motion as an extension of the previous work of \citet{lainey2007first, jacobson2010orbits, jacobson2014martian} and it provides potential opportunity for future use of high-precision observation from future missions to further the study of Phobos and Mars. In the future, it is not only the position and orientation that will feasibly be determined by using this new model, but also 
Phobos' gravity coefficients and $k_2$, for instance, when  combined with observation data from future mission.
Finally, it needs to be pointed out is that our modeling process is based on universal principles and can be applied to a full Mars system (including Deimos) or other planetary systems to confront the continuous improvement of the accuracy of observation data

\begin{acknowledgements}
This research is supported by the National Key Research and Development Program of China (2021YFA0715101, 2022YFF0503202), the National Natural Science Foundation of China (12033009, 12103087, 42241116), the International Partnership Program of Chinese Academy of Sciences (020GJHZ2022034FN), the Yunnan Province Foundation (202201AU070225, 202301AT070328), the Young Talent Project of Yunnan Revitalization Talent Support Program, grant from Key Laboratory of Lunar and Deep Space Exploration, Chinese Academy of Sciences (LDSE202004), and open research fund of state key laboratory of information engineering in surveying, mapping and remote sensing, Wuhan University (21P02). Y. Yang thanks Dr.~Val{\'e}ry Lainey for the careful reviewing and nice suggestions that have been incorporated into and improved the paper and kindly provided the subroutine that compute Phobos' ephemeris from NOE ephemerides, thank the discussion with Dr.~Jinling Li (about computing the variational equations of rotational motion). Phobos' ephemerides files can be downloaded from \url{ftp://ftp.imcce.fr/pub/ephem/NOE/}(NOE-4-2020). 
 \end{acknowledgements}

%
%

\bibliographystyle{bibtex/aa} 
\bibliography{aaref} 

\end{document}